\documentclass[]{mn2e}
\usepackage{epsfig}

\title[The origin of sdB stars] {The Origin of Subdwarf B Stars (I):
the Formation Channels}

\author[Han et al.]
{Z.~Han$^1$\thanks{E-mail: zhanwen@public.km.yn.cn}, Ph.~Podsiadlowski$^2$,
P. F. L. Maxted$^3$, T. R. Marsh$^4$, N. Ivanova$^{2,5}$\\
\it
$^1$ Yunnan Observatory, National Astronomical Observatories, 
the Chinese Academy of Sciences, P.O.Box 110, Kunming, 650011, China\\
$^2$ Department of Astrophysics, University of Oxford, Oxford,
OX1 3RH\\
$^3$ Department of Physics, Keele University, Staffordshire, ST5 5BG\\
$^4$ University of Southampton, Department of Physics {\rm\&} Astronomy,
Highfield, Southampton, S017 1BJ\\
$^5$ Department of Physics {\rm\&} Astronomy, Northwestern University,
Evanston, IL 60208
}

\date{\today}

\begin{document}
\maketitle

\begin{abstract}
Subdwarf B (sdB) stars (and related sdO/sdOB stars) are believed to be
helium core-burning objects with very thin hydrogen-rich envelopes.
In recent years it has become increasingly clear from observational
surveys that a large fraction of these objects are members of binary
systems.  To better understand their formation, we here present the
results of a detailed investigation of the three main binary evolution
channels that can lead to the formation of sdB stars: the common
envelope (CE) ejection channel, the stable Roche lobe overflow (RLOF)
channel and the double helium white dwarfs (WDs) merger channel. The
CE ejection channel leads to the formation of sdB stars in short-period
binaries with typical orbital periods between 0.1 and 10\,d, very thin
hydrogen-rich envelopes and a mass distribution sharply peaked around
$\sim 0.46M_\odot$. On the other hand,
under the assumption that all mass transferred is soon lost, 
the stable RLOF channel produces
sdB stars with similar masses but long orbital periods (400 -- 1500\,d) and
with rather thick hydrogen-rich envelopes.  The merger channel gives
rise to single sdB stars whose hydrogen-rich envelopes are extremely thin
but which have a fairly wide distribution of masses (0.4 --
0.65\,$M_{\odot}$).  We obtained the conditions for the formation of
sdB stars from each of these channels using detailed stellar and
binary evolution calculations where we modelled the detailed evolution
of sdB stars and carried out simplified binary population synthesis
simulations.  The observed period distribution of sdB stars in compact
binaries strongly constrains the CE ejection parameters. The best fits
to the observations are obtained for very efficient CE ejection where
the envelope ionization energy is included, consistent with previous
results. We also present the distribution of sdB stars in the $T_{\rm
eff}$ - $\log g$ diagram, the Hertzsprung-Russell diagram and the
distribution of mass functions.
\end{abstract}

\begin{keywords}
binaries: close -- stars: OB subdwarfs -- stars: white dwarfs
\end{keywords}

\section{Introduction}

Subdwarf B (sdB) stars were originally defined by Sargent and Searle
\shortcite{sar68} as stars with colours corresponding to those of B
stars in which the Balmer lines are abnormally broad compared to those
seen in population I main-sequence stars. Subdwarf O (sdO) stars and
subdwarf OB (sdOB) stars are related stars of correspondingly earlier
spectral type (see, e.g., Vauclair \& Liebert 1987).  Based on an
interpretation of their evolutionary state, sdB stars are also
sometimes referred to as extreme horizontal branch stars. They are
generally considered to be core helium burning stars with extremely
thin hydrogen envelopes ($<0.02M_\odot$), and most of them are
believed to have masses around $0.5M_\odot$
\cite{heb86,saf94}. Indeed, a recent asteroseismological analysis by
Brassard et al.\ (2001) has confirmed a mass of 0.49$\pm
0.02M_{\odot}$ for the sdB star PG 0014+067.  In this paper, we
collectively refer to core helium burning stars with thin hydrogen
envelopes as sdB stars if they are located in the corresponding region
in a ($T_{\rm eff}$, $\log g$) diagram, even if some of
them may in reality be sdO or sdOB stars.

Subdwarf B stars form an important class of objects in several
respects.  At the Galactic level, they are the dominant population in
surveys of blue objects \cite{gre86} and constitute a population of
stars that are important for our understanding of the structure and
evolution of the Galaxy. Pulsating sdB stars \cite{kil99} provide a
standard candle for distance determinations. On a larger cosmological
scale they have been used to constrain the ages of the oldest galaxies
and hence cosmological models. The latter is based on measuring the
age of giant elliptical galaxies from the ultraviolet (UV) excess, or
``upturn'', with the help of evolutionary population synthesis models
where low mass core-helium burning stars provide the dominant source
of UV radiation \cite{bro97,yi97,yi99}.

More importantly, sdB stars are exotic objects because of their thin
hydrogen-rich envelopes. Understanding the process of their formation
helps to improve our understanding of the theory of stellar and binary 
evolution.

There have been extensive surveys of sdB stars in the past.
Magnitude-limited and colour-selected samples have been obtained
from the Palomar Green
(PG) survey \cite{gre86} (magnitude limit $B\sim 16.1$) and the Kitt
Peak Downes (KPD) survey \cite{dow86} (magnitude limit $B=15.3$).  Saffer et
al.\ (1994) measured atmospheric parameters, such as effective
temperature, surface gravity and photospheric helium abundance, for 68
sdB stars.  Ferguson, Green \& Liebert \shortcite{fer84} found 19 sdB
stars with main sequence (MS) companions from the PG
survey and derived a binary frequency of about 50 percent. Allard et
al.\ \shortcite{all94} found 31 sdB binaries from 100 candidates
chosen from the PG and the KPD surveys and estimated that
54 to 66 percent of sdB stars are in binaries with MS companions after
taking selection effects into account.  Thejll, Ulla \& MacDonald
\shortcite{the95} and Ulla \& Thejll \shortcite{ull98} also found that
more than half of their sdB star candidates showed infrared flux
excesses, indicating the presence of binary companions. Aznar Cuadrado
\& Jeffery \shortcite{azn01} obtained atmospheric parameters for 34
sdB stars from spectral energy distributions and found that 19 were
binaries with MS companions, while 15 appeared to be single.  These
observations showed that at least half of the sdB stars were in
binaries.

A major recent development has been the identification of many sdB
stars as short-period 
binaries \cite{saf98,koe98,jef98,woo99,oro99,mor99,max00a,max00b,max01,heb02}.
In particular, Maxted et al.\ \shortcite{max01} concluded that more than two 
thirds of their candidates were binaries with 
short orbital periods from hours to days and that most of the known
companions were white dwarfs (WDs).

A variety of formation channels for sdB stars have been proposed in
the past but mainly for single sdB stars because of the absence of
identified sdB star binaries at the time.  In the merger channel, two
helium white dwarfs in a close binary are driven together by the 
orbital angular momentum loss due to gravitational wave radiation. When
the white dwarfs merge and the merged object ignites helium, this
produces a single sdB star \cite{web84,ibe86,han98}.  Alternatively, stellar
wind mass loss near the tip of the first giant branch (FGB) may strip
off a giant's envelope and leave an almost bare helium core.  If
helium is ignited in the core, the star will appear as a single sdB
star \cite{dcr96}. Sweigart \shortcite{swe97} has studied the
evolution of globular-cluster stars and suggested that helium mixing
driven by internal rotation substantially increases the helium
abundance in the envelope; this may lead to enhanced mass loss along
the FGB and the formation of a sdB star.

On the other hand, Mengel, Norris \& Gross \shortcite{men76} carried
out conservative binary evolution calculations for a binary system
with initial masses of $0.80M_\odot$ and $0.78M_\odot$ and a
composition $X=0.73$, $Z=0.001$, and showed that there exists a range
of initial separations for which stable mass transfer can produce an
sdB star of $\sim 0.5M_\odot$ in a wide binary.

From a binary evolution point of view, these formation channels are
not complete.  When a star fills its Roche lobe near the tip of the
FGB, mass transfer begins and may be dynamically unstable.  This leads
to the formation of a common envelope (CE) \cite{pac76}, where the CE
engulfs the helium core and the secondary.  Due to friction between
the envelope and the immersed binary, the orbit shrinks, depositing a
large amount of orbital energy in the envelope. If this energy is
enough to eject the envelope and if helium is subsequently ignited in
the core, a sdB star in a short-period binary is formed with a mass
near $0.5M_\odot$. These are exactly the types of objects identified
in large numbers by Maxted et al.\ \shortcite{max01}.  If mass
transfer near the tip of the FGB is dynamically stable, the envelope
of the primary is lost as a result of stable RLOF, and the remnant
core will be in a binary system with a long orbital period. It becomes
a sdB star when helium in the primary's remnant is ignited. An
additional channel for the formation of sdB stars in wide binaries,
which has not received much attention in the past, involves binaries
that experience stable RLOF when passing through the Hertzsprung gap
(so-called early case B mass transfer) (Han, Tout \& Eggleton 2000;
Han et al. 2002; in preparation [henceforth, Paper II]).  All of the
sdB binaries produced through stable RLOF channels are consistent with
the observations by Green, Liebert \& Saffer \shortcite{gre00} who
showed that some sdB stars appear to be members of long-period
binaries.

The main purpose of this study is to re-examine the various scenarios
for the formation of sdB stars in some detail. In this first paper, we
concentrate on the individual evolutionary channels. Using detailed
stellar and binary calculations, we model the physics and appearance
of sdB stars and then test individual evolutionary channels using
binary population synthesis (BPS). We demonstrate that all of the main
evolutionary channels proposed previously can lead to the formation of
sdB stars. As a by-product we constrain the CE ejection efficiency
from the observed period distribution of compact sdB binaries to
arrive at a physically motivated and experimentally calibrated
prescription for the CE phase.

The outline of this paper is as follows.  In section 2, we describe
the stellar evolution code and the binary population synthesis code
adopted in this study. In section 3, we present the conditions for the
formation of sdB stars from the CE ejection channel, their
evolutionary tracks and simplified BPS models to constrain the CE
ejection efficiency.  In section 4, we derive the conditions for
helium ignition in objects that result from the merger of two He white
dwarfs and use Monte Carlo simulations to determine their mass
distribution.  In section 5, we investigate the criterion for stable
RLOF and the formation of sdB stars in wide binaries.  In the follow-up
paper (Paper II) we will apply these results to a comprehensive binary
population synthesis study and will estimate the relative importance
of these individual channels.

\section{The stellar evolution and the binary population synthesis code}

In this study, we employ two numerical computer codes: a stellar
evolution code to determine the structure and follow the evolution of
sdB stars, and a binary population synthesis code to examine different
evolutionary channels.

The stellar evolution code used is the one that has been originally developed
by P. P. Eggleton \shortcite{egg71,egg72,egg73},
which has been updated with the latest
input physics over the last 3 decades as
described by Han, Podsiadlowski \& Eggleton\ 
\shortcite{han94} and  Pols et al.\ \shortcite{pol95,pol98}.
The code distinguishes itself by the use of a
self-adaptive non-Lagrangian mesh, the treatment of both
convective and semiconvective mixing as diffusion processes  and the
simultaneous and implicit solution of both the stellar structure
equations and the chemical composition equations which includes
convective mixing. These characteristics
make the code very stable and easy to use.
The current code uses an equation of state that includes
pressure ionization and Coulomb interaction, recent
opacity tables derived from Rogers \& Iglesias \shortcite{rog92}
and Alexander \& Ferguson \shortcite{ale94a,ale94b}, nuclear reaction rates
from Caughlan \& Fowler \shortcite{cau88} and Caughlan et al.\
\shortcite{cau85}, and neutrino loss rates from Itoh et al.\
\shortcite{ito89,ito92}.

We set $\alpha =l/H_{\rm p}$, the ratio of the mixing length
to the local pressure scale height, to 2. 
For Population I (Pop I) stars (with a typical composition
of hydrogen abundance $X=0.70$, helium abundance $Y=0.28$ and
metallicity $Z=0.02$), such a value for $\alpha$
gives a roughly correct lower main sequence, as determined observationally
by Andersen \shortcite{and91}.  It also well reproduces the location of the
red giant branch in the HR diagram for stars in the Hyades supercluster
\cite{egg85}, as determined by Bessell et al.\ \shortcite{bes89}.
A fit to the Sun also leads to $\alpha =2$ as the most
appropriate choice \cite{pol98}.

For convective overshooting, the code does not employ a prescription
in terms of the pressure scaleheight $H_{\rm p}$ where the
overshooting length is a fixed fraction of $H_{\rm p}$.  Instead the
code uses an approach based on the stability criterion itself, the
`$\delta_{\rm ov}$ prescription', by incorporating a condition that
mixing occurs in a region with $\nabla_{\rm r}>\nabla_{\rm a} -
\delta_{\rm ov}/(2.5+20\zeta +16\zeta ^2)$, where $\zeta$ is the ratio
of radiation pressure to gas pressure and $\delta_{\rm ov}$ is a
specified constant, the overshooting model parameter.  Critical tests
of stellar evolution by means of double-lined eclipsing binaries
\cite{sch97,pol97} show that $\delta_{\rm ov}=0.12$ gives the best fit
to the observed systems, where $\delta_{\rm ov}=0.12$ corresponds to
an overshooting length of $\sim 0.25$ pressure scale heights.

Roche lobe overflow is treated directly within the code. It has been
tested thoroughly and works very reliably. Because the mesh-spacing
is computed along with the structure, the inclusion of
RLOF is almost trivial: it just requires a modification of one 
surface boundary condition.
The boundary condition is written as
\begin{equation}
{{\rm d}m\over {\rm d}t}=C\cdot {\rm Max}\left[0,
({r_{\rm star}\over r_{\rm lobe}}-1)^3\right]
\label{boundary}
\end{equation}
where ${\rm d}m/ {\rm d}t$ gives the rate at which the mass
of the star changes,
$r_{\rm star}$ is the radius of the star, and $r_{\rm lobe}$ the
radius of its Roche lobe. $C$ is  a constant.
With $C=1000 M_\odot/$yr, RLOF proceeds steadily, and
the lobe-filling star
overfills its Roche lobe as necessary
but never overfills its lobe by a substantial amount (typically
$({r_{\rm star}/ r_{\rm lobe}}-1)\la 0.001$).

The stellar evolution code described above evolves only a single star
or both components of a binary at a time. However, stellar evolution
theory should give and predict the {\it statistical\/} properties of a
whole stellar population as well as the properties of individual stars
or binaries.  In order to investigate statistical properties of stars
and check evolutionary mechanisms for different types of stars, Han,
Podsiadlowski \& Eggleton \shortcite{han95c} developed a Monte Carlo
simulation code, or `binary population synthesis' (BPS) code, which is
able to evolve a sample of 1 million or more stars (including
binaries) simultaneously by interpolating the properties of individual
stars as a function of evolutionary age in a specially prepared grid
of stellar models. This code has been steadily updated ever since
\cite{han95a,han95b,han98,han01}.

The BPS code needs a grid of stellar evolution models.  Taking Pop I as the
standard population model, we carried out a large number of 
stellar evolution calculation for a wide range 
of masses ($0.08M_\odot$ to $126M_\odot$), including
the evolution of helium stars from $0.32M_\odot$ to $8M_\odot$,  at an
interval of $\sim 0.1$ in $\log M$. In the
calculation of the stellar models, we did not include stellar wind mass
loss for most of the grids calculated.
However, for massive stars the evolution was terminated at the
Humphreys-Davidson (HD) limit \cite{hum79,lam88,fit90,ulm98}; we assumed
that at the HD limit the envelope of the massive star was lost completely,
and then treated the remnant core as a helium star. The evolution 
of low- and intermediate-mass stars was
terminated at the point when the total energy of the envelope became 
positive, assuming that the envelope was ejected at this point, leaving
a white dwarf remnant \cite{han94}. For a given star we obtain
the required stellar parameters at a particular evolutionary age
(e.g. the luminosity, effective temperature, radius, core mass, core radius,
envelope binding energy) by interpolation from our set of stellar evolutionary 
tracks.

In BPS, one has to evolve binaries as well as single stars. At present
stellar wind mass loss and binary interactions are included
in the form of simple prescriptions.  We plan to replace these by ever more
realistic modelling as part of our ongoing work. One important
improvement we have implemented in the present work is that we adopted
full binary evolution calculations for systems where RLOF occurs in
the Hertzsprung gap \cite{han00}.  Another uncertainty which we will
address in more detail in the future is the criterion for dynamically unstable
RLOF. If the primary fills its Roche lobe as a red giant, RLOF may be
dynamically unstable if the mass ratio at the onset is larger than
some critical value $q_{\rm c}$, given e.g. by Hjellming \& Webbink
\shortcite{hje87} and Webbink \shortcite{web88}. However, this
critical mass ratio only applies to conservative RLOF and does not
fully take into account the detailed dynamics at the onset of mass
transfer.  From observations it is clear that RLOF on the FGB/AGB is 
non-conservative \cite{gia81,sho88}.
Part of the transferred mass is lost from the
system. In our model we assume that the lost matter carries away the same
specific angular momentum as pertains to the system.  Defining a
mass transfer efficiency, $\alpha_{\rm RLOF}$, as the ratio of the mass
accreted by the secondary to the mass transferred from the primary
($\alpha_{\rm RLOF}=1$ for conservative RLOF), we find that the
critical ratio $q_{\rm c}$ depends strongly on the mass transfer
efficiency \cite{han01}.  Dynamically unstable mass transfer may
result in the formation of a common envelope (CE) \cite{pac76},
leading either to the formation of a close binary or the complete
merger of the two components (see section 3).
 
Given a binary sample, the BPS code performs all the necessary
interpolations in the model grid, integrates the mass loss along
evolutionary tracks for an assumed stellar-wind law and deals with all
the binary interactions. The output of the code are all 
the parameters for different types of binaries or single objects formed as
a consequence of the evolution and various interactions.

In the original version, the BPS code used the following stellar model grids 
as input:
\begin{enumerate}
\item 
  Z=0.02, no stellar wind, no overshooting, 0.08-126.0$M_\odot$\ 
  for normal stars ($X=0.70$, $Y=0.28$), 0.32-8.0$M_\odot$\ for helium stars.
\item
  Z=0.004, no stellar wind, no overshooting, 0.1-126.0$M_\odot$\ 
  for normal stars ($X=0.74$, $Y=0.256$), 0.32-8.0$M_\odot$\ for helium stars.
\item
  Z=0.001, no stellar wind, no overshooting, 0.1-126.0$M_\odot$\ 
  for normal stars ($X=0.75$, $Y=0.249$), 0.32-8.0$M_\odot$\ for helium stars.
\end{enumerate}
For the purpose of this paper, we have also calculated several additional grids
that include convective overshooting and stellar wind mass loss:
\begin{enumerate}
\item
  Z=0.02, no stellar wind but with overshooting, 0.63-3.2$M_\odot$ 
  for normal stars.
\item
  Z=0.02, 1/4 of Reimers' wind \cite{rei75}
  and with overshooting, 0.63-3.2$M_\odot$
  for normal stars.
\item
  Z=0.02, 1/4 of Reimers' wind but without overshooting, 0.63-3.2$M_\odot$
  for normal stars.
\item
  Z=0.02, 1/2 of Reimers' wind and with overshooting, 0.63-3.2$M_\odot$
  for normal stars.
\item 
  Z=0.02, 1/2 of Reimers' wind but without overshooting, 0.63-3.2$M_\odot$
  for normal stars.
\item
  Z=0.004, 1/4 of Reimers' wind and with overshooting, 0.63-3.2$M_\odot$
  for normal stars.
\end{enumerate}

\section{The common-envelope channel}

In the common-envelope (CE) channel, the sdB star forms in a close
binary as a consequence of dynamical mass transfer and a CE phase
where the progenitor, a giant star, starts to fill its Roche lobe when
it is relatively close to the tip of the first red-giant branch
(FGB). This situation generally occurs when the radius of the
mass-losing star increases faster than its Roche-lobe radius. This
leads to mass transfer on a dynamical timescale and the formation of a
common envelope where the envelope of the giant engulfs both its
degenerate core and the companion star.  Friction between these
orbiting components and the envelope causes the orbit of the immersed
binary to shrink. If the orbital energy released in the process is
able to eject the envelope, this process leaves a very tight binary
consisting of the degenerate core of the giant and the companion
star. This is believed to be the main mechanism, originally proposed
by Paczy\'nski (1976), by which an initially wide binary is
transformed into a very close system. If this happens when the giant
was sufficiently close to the tip of the FGB at the beginning of mass
transfer, i.e. the core was close to experiencing the helium flash,
the remnant core of the giant may still ignite helium (as first
demonstrated by Castellani \& Castellani 1993) and hence become a
helium core-burning sdB star, where the companion can be either a
white dwarf (WD) or in some cases a low-mass star. These are exactly
the objects observed in large numbers by Maxted et al.\
\shortcite{max01}. Note that the companion star can in principle also
be a normal dwarf star with a mass as high as $1-2M_\odot$ (which
depends on the condition for dynamical mass transfer; see \S
5.1). However, such a system would have a composite spectrum. Since
such systems were excluded from the PG catalog, they would not
appear in radial-velocity studies based on this catalog.

The details of CE evolution, in particular the conditions for which
the envelope can be ejected are far from well understood at the
present time (see e.g. Iben \& Livio 1993). On the other hand, the
identification of a large number of sdB stars in very close binaries,
which must all have passed through such a well-defined evolutionary
channel, provides a unique opportunity to test particular models of CE
evolution and may even help to calibrate the criterion for CE
ejection. Indeed, it is one of the surprises of the observations by
Maxted et al.\ \shortcite{max01} 
that the period distribution is extremely
wide ranging from 2\,hr to more than $\sim 10\,$d.

\subsection{The minimum core mass for helium ignition}

\begin{table}
 \caption{The minimum core mass for the helium flash/helium ignition}
 \begin{tabular}{lllll}
 \hline\hline
 $M_0$ & $M_{\rm c}^{\rm min}$ & $M_{\rm c}^{\rm tip}$ &
         $\log ({R^{\rm min}\over R_\odot})$ 
         & $\log ({R^{\rm tip}\over R_\odot})$ \\
 ($M_\odot$) &  ($M_\odot$) &  ($M_\odot$) \\
 \hline
 &&&&\\
 \multicolumn{5}{l}{No wind, no overshooting, $Z=0.02$}\\
  0.794 &  0.4529 &  0.4742 &  2.2245 &  2.2923 \\
  1.000 &  0.4502 &  0.4725 &  2.1838 &  2.2539 \\
  1.259 &  0.4492 &  0.4720 &  2.1492 &  2.2161 \\
  1.585 &  0.4461 &  0.4699 &  2.1069 &  2.1772 \\
  1.995 &  0.4174 &  0.4422 &  1.9537 &  2.0453 \\
  2.114 &  0.4010 &  0.4217 &  1.8628 &  1.9532 \\
  &&&&\\
 \multicolumn{5}{l}{Reimers' wind ($\eta=1/4$), with overshooting, $Z=0.02$}\\
  0.800 &  0.4538 &  0.4746 &  2.2482 &  2.3076 \\
  1.000 &  0.4509 &  0.4727 &  2.1985 &  2.2707 \\
  1.260 &  0.4496 &  0.4723 &  2.1578 &  2.2264 \\
  1.600 &  0.4351 &  0.4601 &  2.0656 &  2.1465 \\
  1.900 &  0.3889 &  0.4087 &  1.8113 &  1.9050 \\
 \multicolumn{5}{l}{Helium ignited non-degenerately}\\
  2.050 &  0.3190 &  0.3224 &  1.4290 &  1.4355 \\
  2.265 &  0.3181 &  0.3460 &  1.2881 &  1.5317 \\
  2.512 &  0.3223 &  0.3715 &  0.6871 &  1.5612 \\
  3.162 &  0.4385 &  0.4669 &  0.7489 &  1.6891 \\
  3.981 &  0.5947 &  0.6146 &  0.8068 &  1.8685 \\
  5.012 &  0.8049 &  0.8286 &  0.8598 &  2.0618 \\
  6.310 &  1.1108 &  1.1284 &  0.9181 &  2.2587 \\
  &&&&\\
 \multicolumn{5}{l}{Reimers' wind ($\eta=1/4$), with overshooting, $Z=0.004$}\\
  0.800 &  0.4642 &  0.4855 &  2.0886 &  2.1475 \\
  1.000 &  0.4604 &  0.4826 &  2.0530 &  2.1184 \\  
  1.260 &  0.4578 &  0.4810 &  2.0186 &  2.0856 \\
  1.600 &  0.4314 &  0.4564 &  1.8829 &  1.9710 \\
  1.750 &  0.3933 &  0.4149 &  1.6800 &  1.7825 \\
 \multicolumn{5}{l}{Helium ignited non-degenerately}\\
  1.850 &  0.3238 &  0.3367 &  1.4003 &  1.4439 \\
  2.000 &  0.3175 &  0.3403 &  1.2903 &  1.4594 \\
  2.512 &  0.3381 &  0.3935 &  0.5740 &  1.5122 \\
  3.162 &  0.4492 &  0.4942 &  0.6343 &  1.6480 \\
  3.981 &  0.6006 &  0.6398 &  0.6910 &  1.8080 \\
  5.012 &  0.8143 &  0.8579 &  0.7515 &  1.9889 \\
  6.310 &  1.1187 &  1.1539 &  0.8126 &  2.1707 \\
 \hline
 \end{tabular}

 \medskip
  Note - $M_0$: initial zero-age main-sequence mass;
  $M_{\rm c}^{\rm min}$: minimum mass of the He core for helium ignition;
  $M_{\rm c}^{\rm tip}$: mass of the He core at the tip of the first red-giant
  branch;
  $R^{\rm min}$: stellar radius corresponding to $M_{\rm c}^{\rm min}$;
  $R^{\rm tip}$: the radius at the tip of the first red-giant branch.
  The models were calculated by taking off mass from the envelope
  at a rate of $10^{-3}M_\odot {\rm yr}^{-1}\times$ the mass of the star 
  in solar units
 until the envelope collapsed.
  Note, however, that $M_{\rm c}^{\rm min}$ and $R^{\rm min}$ for 
  $M_0\geq 2.5M_\odot$ correspond to models at the end of the main sequence,
  while those with  $M_0 < 2.5M_\odot$ are models on the FGB. 
  \label{min-core-tab}
\end{table}

\begin{figure}
\epsfig{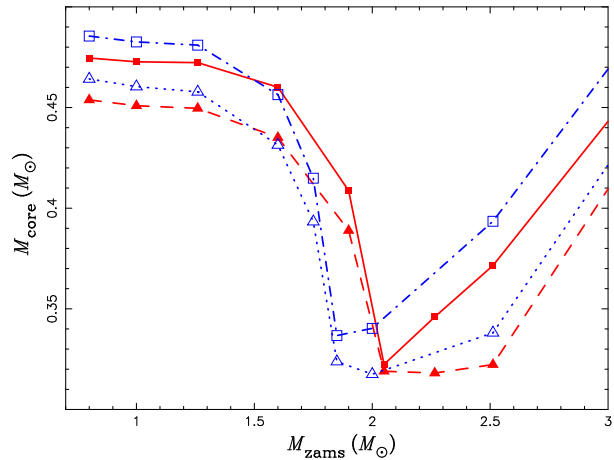}
\caption{Range of core masses for the occurrence of a He flash (or 
 non-degenerate helium ignition)
 as a function of initial mass for $Z=0.02$ (dashed and solid curves) and
 $Z=0.004$ (dotted and dot-dashed curves). The lower curve
 for each set gives the minimum core mass above which a star burns helium,
 the upper curve gives the core mass at the normal tip
 of the first giant branch (FGB)}
\label{min-core}
\end{figure}
In order for the degenerate core to ignite helium in a helium flash
after the envelope has been ejected, its giant progenitor had to be
relatively close to the tip of the FGB at the beginning of mass
transfer. D'Cruz et al.\ \shortcite{dcr96} have shown by varying the wind
mass-loss rate on the FGB that the core mass for helium ignition can
be as low as $0.45M_\odot$. The situation considered here is slightly different
since mass ejection in a CE phase is a sudden event, occurring on a
timescale short compared to the evolutionary timescale of a giant.  We
therefore performed a comprehensive series of stellar calculations
where we assumed that a giant loses its hydrogen-rich envelope near
the tip of the FGB to determine the minimum core mass above which
the core will
still ignite helium after the ejection of the envelope. This critical
mass generally depends on the ZAMS mass of the giant, but also on the
assumptions about mass loss, metallicity and the degree of convective
overshooting from the core. To examine these various situations, we
therefore considered three sets of calculations.  Set (1) assumes no
stellar wind, no convective overshooting and a solar metallicity of
$Z=0.02$. In set (2) we again use a solar metallicity, but also include
a stellar wind parametrized by a Reimers' wind mass-loss law
\begin{equation}
\dot{M}_{\rm wind}= 4\times 10^{-13} \eta RL/M,
\label{reimers}
\end{equation}
where for the standard model we use an efficiency $\eta =1/4$
\cite{ren81,ibe83,car96}.
In this set, we also take into
account convective overshooting based on the calibration of this
parameter by Schr\"oder et al.\ \shortcite{sch97} and 
Pols et al.\ \shortcite{pol97}, which
corresponds to overshooting of $\sim 0.25$ pressure scale heights from
the core.  Set (3) is similar to set (2) except that we use a
metallicity of $Z=0.004$, characteristic of a thick disc population
\cite{gil89}.

We find that the maximum initial ZAMS mass below which stars
experience a helium flash decreases from $2.25M_\odot$ for set (1) to
$1.99M_\odot$ for set (2) and $1.8M_\odot$ in set (3).  Stars with ZAMS
masses larger than these values ignite helium non-degenerately. Some
of these can also become sdB stars. 

To determine the minimum core mass for the helium flash for each ZAMS
mass, we considered a series of models near the tip of the FGB and
took the H-rich envelopes off at a high rate (chosen to be
$10^{-3}M_\odot {\rm yr}^{-1} \times$ 
the mass of the star in solar units) until the
envelopes collapsed (note that we switched hydrogen burning off when
the envelope mass became less than $0.002M_\odot$ to prevent the occurrence
of hydrogen shell flashes). We then followed the subsequent evolution
of the core either until it had cooled to a surface temperature of less than
$5000\,$K
or until it ignited
helium in the core. In Table~\ref{min-core-tab} 
and Figure~\ref{min-core} we present the minimum
core mass as a function of ZAMS mass above which helium is ignited and
also the core mass at the tip of the FGB. These results show that the
minimum core mass has to be typically within 5 per cent of the core
mass at the tip of the FGB where the minimum decreases from
0.45/0.46$M_\odot$ for the less massive progenitors to 0.39/0.40$M_\odot$
for the most massive stars that still experience a helium flash. While
the range in mass is relatively small, it corresponds to a fairly
large range ($\sim 15$ per cent)
in radius (also shown in Table~\ref{min-core-tab}), 
since giants expand quite significantly close to the tip of
the FGB.

Note that the more massive stars do not experience a helium flash.
If the envelopes are stripped off near the tip of the FGB for
stars with $2.05\leq M_0\leq 2.265M_\odot$ (set 2, Pop I) or with
$1.85\leq M_0\leq 2.0M_\odot$ (set 3, $Z=0.04$), 
the cores ignite helium under non-degenerate conditions (as shown
in Table 1 and Figure 1).
For even more massive stars, the cores will burn helium even when
the envelopes are lost in the Hertzsprung gap.
The envelopes of these massive stars are generally much more tightly bound 
than those near the tip of the FGB for stars with degenerate cores
and hence are much harder to eject. Generally, the more massive the star, the
more tightly bound is its envelope. Therefore we find in the BPS
calculations that CE ejection is only possible for stars that are 
not too massive. These, however, leave fairly low-mass sdB
stars ($\sim 0.35M_\odot$) with very short orbital periods.

\subsection{Evolutionary tracks of sdB stars}

\begin{figure*}
\epsfig{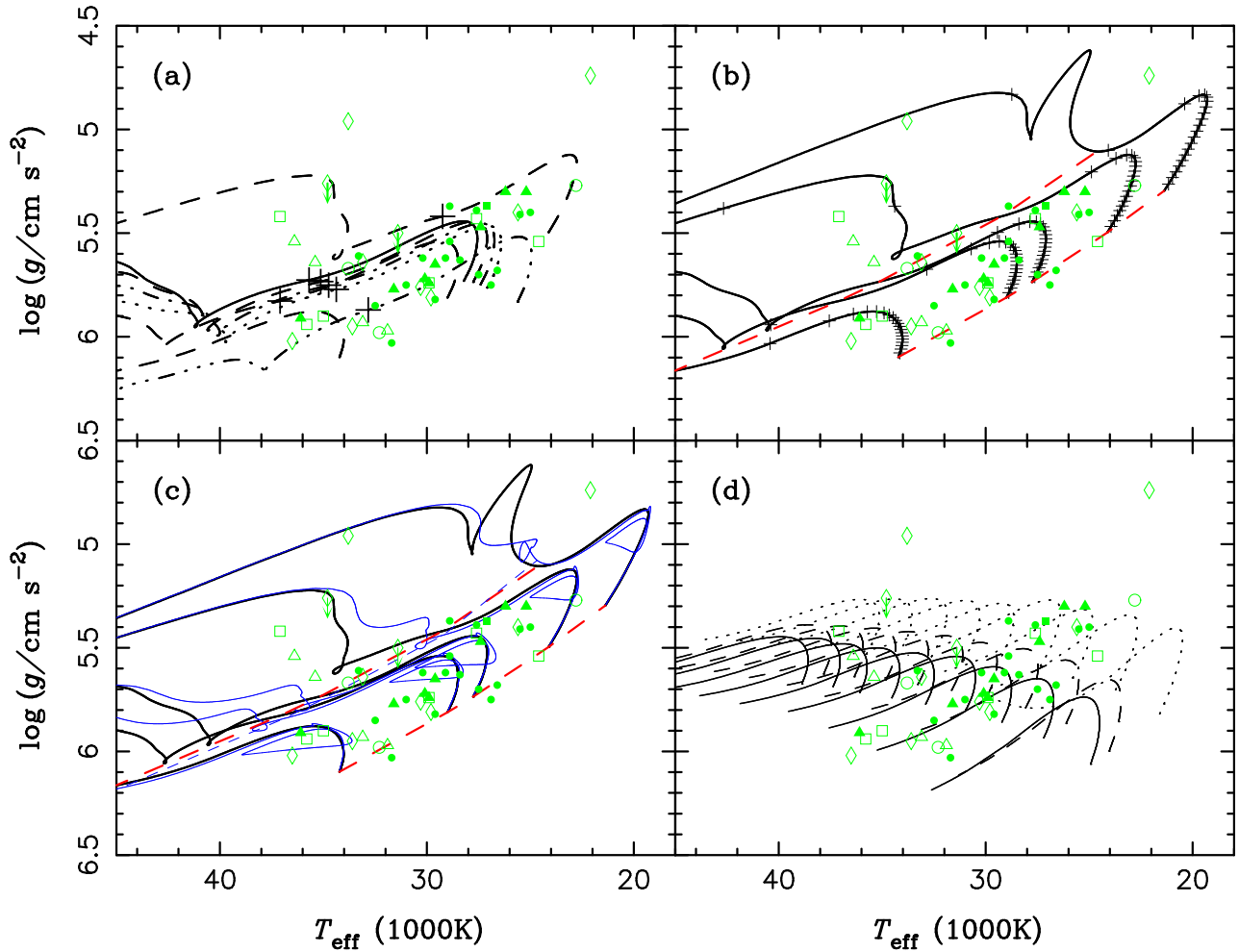}
\caption{
Evolutionary tracks of sdB stars in the $T_{\rm eff}$ - $\log g$
diagram.  Filled circles show
the position of observed sdB stars with orbital periods
$P_{\rm orb}<1{\rm d}$, solid triangles for systems with period
$1<P_{\rm orb}<10{\rm d}$, solid squares are for systems with
$P_{\rm orb}>10{\rm d}$. Circles show systems that have radial
velocity variations ${\rm d}V>40{\rm km/s}$, triangles are for systems with
$20<{\rm d}V<40{\rm km/s}$, squares for $10<{\rm d}V<20{\rm km/s}$, 
diamonds for ${\rm d}V<10{\rm km/s}$, where ${\rm d}V$ is the maximum 
difference between radial velocities measured for a particular object.
Arrows indicate lower limits for $g$.
Panel (a): tracks for 8 selected models (Pop I, Reimers' wind with
$\eta = 1/4$ and convective overshooting). The solid curve is for a ZAMS model
of $0.8M_\odot$ and a sdB of $0.47M_\odot$ with an envelope mass of 
$0.002M_\odot$. The dashed curves are for a ZAMS model of
$1.00M_\odot$ and a sdB mass of $0.46M_\odot$ but
with decreasing envelope masses (top to bottom: 0.005, 0.002, 0.001 and 
$0.000M_\odot$, respectively). The dot-dashed curve is for a ZAMS mass of
$1.26M_\odot$ and a sdB mass of $0.45M_\odot$ and an envelope mass of 
$0.002M_\odot$. The dotted curve is for a ZAMS mass of 
$1.60M_\odot$ and a sdB mass of $0.44M_\odot$ and an
envelope mass of $0.002M_\odot$.
The dot-dot-dot-dashed curve is for a ZAMS mass of 
$1.90M_\odot$ and a sdB mass of $0.40M_\odot$ and an
envelope mass of $0.002M_\odot$.
Crosses show the point of central He exhaustion. 
Panel (b) illustrates the dependence of the evolutionary tracks on the envelope
mass. All models are for a ZAMS model of $1M_\odot$ and a sdB mass of
$0.46M_\odot$ (for Pop I, Reimers' wind with $\eta=1/4$ and with convective
overshooting). The solid curves from bottom to top are
for envelope masses of 0.000, 0.001, 0.002, 0.005, $0.010M_\odot$, 
respectively. The left dashed curve indicates the point of
central helium exhaustion, while the right dashed curve shows the locus
of zero-age HB models. The age differences between adjacent crosses
are $10^7{\rm yr}$. 
Panel (c) illustrates the dependence of evolutionary tracks on convective
overshooting. The thin solid/dashed curves do not include convective 
overshooting, while the solid ones do (the latter are the same as in 
panel (b)). 
Panel (d) illustrates the variation with sdB mass (for $Z=0.02$, with 
overshooting). Solid curves are for an envelope mass of $0.001M_\odot$, dashed
curves for $0.002M_\odot$ and dotted curves for $0.005M_\odot$. For each
set, the curves from right to left are for sdB masses of 0.35, 0.40, 0.45,
0.50, 0.55, 0.60, 0.65, 0.70, $0.75M_\odot$, respectively. 
All curves show the tracks from the zero-age HB to the point of 
central helium exhaustion. 
}
\label{sdb-evol}
\end{figure*}

\begin{figure}
\epsfig{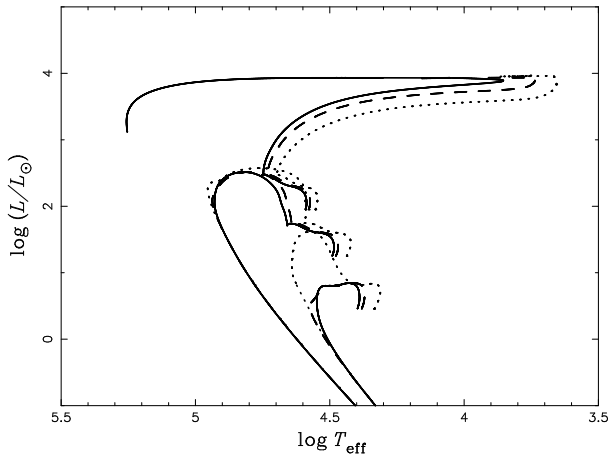}
\caption{
H-R diagram of sdB stars. Solid curves are for envelope masses
of $0.001M_\odot$, dashed curves for $0.002M_\odot$ and dotted
curves for $0.005M_\odot$, respectively. For each set, the curves from
bottom to top are for sdB masses of 0.35, 0.50, $0.75M_\odot$,
respectively. Note that the composition profiles were not
taken from helium flash models, but were constructed from
a $3.2M_\odot$ Pop I star burning helium smoothly. The sdB model
for the $0.75M_\odot$ star was terminated when it experienced
a hydrogen shell flash.
}
\label{sdb-hrd}
\end{figure}

Since helium ignition in a degenerate core causes a mild thermonuclear
runaway which leads to the expansion of the core on a dynamical
timescale, we are unable to follow it with our hydrostatic
stellar-evolution code. However, once the core has expanded
sufficiently and the core has become non-degenerate, helium burning
quickly stabilizes and the core regains hydrostatic equilibrium. In
Figures~\ref{sdb-evol} and~\ref{sdb-hrd} we show the subsequent evolution of 
various sdB stars in
a $T_{\rm eff}$ -- $\log g$ diagram and a standard Hertzsprung-Russell
(H-R) diagram. The $T_{\rm eff}$ -- $\log g$ is particularly useful,
since sdB stars can be placed on it based on their spectra alone with
the help of model stellar atmospheres independent of
their distance and their luminosity. The initial models
(i.e. on the zero-age horizontal branch) in panels (a), (b) and
(c) of Figure~\ref{sdb-evol}
were constructed in such a way that they had the same core mass and
chemical composition as the appropriate models from set (2) in \S~3.1
at the point of helium ignition (i.e. at the helium
flash). Panel (a) shows the evolution of various selected sdB models
with different total masses, originating from a range of progenitor
masses and with a variety of different envelope masses. The symbols
give the position of observed sdB stars (from Maxted et al.\
2001; note, however, the uncertainties in $\log g$ are about 0.15;
see Saffer et al.\ 1994), 
where the different symbols correspond either to different
ranges of orbital periods (if known; filled symbols) or indicate the
magnitude of radial velocity variations for systems without known
orbital periods (open symbols).  As the figure shows, there
is generally excellent overlap between the evolutionary tracks and the
observed systems. To show this more systematically, panel (b)
of Figure~\ref{sdb-evol} presents
the tracks for sdB stars of a typical mass of $0.46M_\odot$, but with
different masses of the hydrogen-rich envelope, ranging from 0 to
$0.01M_\odot$ 
(here the envelope is defined as the outer part of the sdB star with 
a hydrogen mass fraction larger than 0.01).
The two dashed
curves give the location of the ZAHB (right curve) and the location
where the sdB stars exhaust helium in their centres (left curve). Tick
marks are also shown along the tracks to indicate the speed of
evolution across the diagram. As one would expect, sdB stars without
hydrogen envelopes are the hottest and the most compact, and the
evolutionary tracks are shifted towards lower temperatures and lower
gravity as the mass in the hydrogen-rich envelope is increased. The
majority of observed systems lie between the dashed curves, marking
the range of helium core-burning objects.  There appears to be,
however, a fairly large range of envelope masses from 0 to $\sim
0.005M_\odot$. There are also a few objects that lie significantly
outside the helium core-burning band. While evolutionary tracks will
pass through most of these observational points after the end of
helium core burning, this corresponds to a fast evolutionary phase, as
can be seen from the wide separation of tick marks. A more likely
explanation is that some of these are more massive objects. In panel
(c) of Figure~\ref{sdb-evol} we show the dependence of evolutionary
tracks of sdB stars on convective overshooting. The sdB stars without
convective overshooting display `breathing pulses' (see e.g.
Castellani et al.\ 1985).  In panel (d) of Figure~\ref{sdb-evol}
 we show the dependence of the location of the helium
core-burning sdB stars on the mass of the sdB star for 3 envelope
masses (0.001, 0.002, $0.005M_\odot$). The masses of the sdB stars range
from 0.35 to $0.75M_\odot$ (note that sdB stars more massive than $\sim
0.5M_\odot$ can form in some of the other evolutionary channels
discussed later in this paper)\footnote{ Unlike the previous models,
the composition profiles for the ZAHB models were taken from a
$3.2M_\odot$ helium core-burning star that was artificially transformed
into a sdB star of the chosen total mass and envelope mass.}.  Four of
the six objects that were outside the helium core-burning band in
panel (b) fall onto a helium core burning track with a more massive
sdB star. This may provide an indication that some of the sdB stars
have masses as high as $0.7M_\odot$. The two remaining objects
(PG 1553+273 and PG 1051+501), 
both of which show low radial velocity variations, lie well
above all of these tracks. This suggests that they have either
substantially more massive envelopes (see section 5)
or have already exhausted helium in their cores and are now
evolving quickly away from the horizontal branch.

\subsection{Common-envelope ejection}
CE evolution is one of the most important but also one of the least
understood phases of binary evolution (see Iben \& Livio 1993;
Taam \& Sandquist 2000; Podsiadlowski 2001 for reviews with different
emphasis). One of the uncertainties is related to the conditions under
which a binary experiences dynamical mass transfer and a CE phase
(this will be further discussed in \S~5.1). A second area of uncertainty
is related to the criterion for the ejection of the CE, which 
crucially determines the orbital period distribution of post-CE binaries. 
The latter depends
on what fraction of the orbital energy that is released in the spiral-in
process can be used to drive the ejection, which depends on the efficiency
with which energy can be transported to the stellar surface where it can
be radiated away. It also depends on the efficiency of the dynamics of
the ejection process; e.g. if the envelope is ejected with a velocity
much larger than the surface escape velocity, the efficiency per unit
mass will be reduced (see e.g. the discussion in Taam \& Sandquist 2000).
A further factor is related to the question what fraction of the thermal
energy, in particular the ionization energy, can be converted into kinetic 
energy and can help to drive the expansion of the envelope. This is
particularly important for giants that fill their Roche lobes near the tip
of the FGB or the AGB since in such extended stars the total binding
energy of the envelope, including the ionization energy, is greatly
reduced and ultimately becomes 0 (see Han, Podsiadlowski \& Eggleton
1994; HPE). This has the consequence that very little spiral-in is required
to release, in principle, enough energy to eject these loosely bound envelopes;
this leads to post-CE binaries with relatively long orbital periods. 

In binary population synthesis (BPS) studies it is commonly assumed
that the common envelope is ejected when the change in orbital energy
times some efficiency factor, $\alpha_{\rm CE}$, exceeds the binding
energy of the envelope, where the latter is often approximated by a
simple analytical expression.  Our approach is rather different from
this and it is designed to provide a more physical parametrization of the
CE-ejection process. Our common-envelope ejection criterion can be
written as
\begin{equation}
\alpha_{\rm CE}\,|\Delta E_{\rm orb}| > |E_{\rm gr} + \alpha_{\rm th}
\,E_{\rm th}|.
\end{equation}
The left-hand side represents the fraction of the change in the
orbital energy that can be used for the ejection, as in most other
commonly used prescriptions. However, on the right-hand side, we include both
the gravitational energy of the envelope ($E_{\rm gr}$) and a fraction
$\alpha_{\rm th}$ of its thermal energy ($E_{\rm th}$), which in
particular includes the ionization energy.  Moreover, instead of using
analytical approximations for these energies, we use the values
obtained from full stellar structure calculations (see HPE for details
and also Dewi \& Tauris 2000). The fact that we have 2 parameters,
$\alpha_{\rm CE}$ and $\alpha_{\rm th}$, instead of one allows us
to specifically assess the importance of the thermal energy contribution.
Indeed, the observations of post-CE sdB binaries from this well-defined
evolutionary channel may allow the calibration of this parameter -- at least
in principle.

\subsection{A simplified BPS model}
In order to test the CE ejection criterion, we need to perform a binary
population study where we simulate the period distribution of sdB binaries
after the ejection of the common envelope. To avoid unnecessary complications,
we use a simplified BPS model in this section, simplified in the sense 
that we do not model 
the evolution of the system before the CE phase that leads to the formation
of the close sdB binary. We also restrict ourselves to systems where
the companion star is likely to be a white dwarf (WD). The model is completely
specified by three distributions: the distribution of the white-dwarf mass,
the mass of the giant and the orbital separation before the CE phase.
We use a uniform distribution in $\log (a/R_\odot )$ from 1 to 4 where
$a$ is the orbital separation, a simple WD mass distribution
($f(M_{\rm WD})=10/3\,M_\odot^{-1}$ for $0.25<M_{\rm WD}/M_\odot< 0.45$ or
$0.55<M_{\rm WD}/M_\odot<0.65$) and a Miller \& Scalo \shortcite{mil79}
mass distribution 
for the giant between 0.8 and $8.M_\odot$. 
The WD mass distribution may seem a little bit odd at first sight, but
it actually mimics the bimodal mass distribution of 
WDs after the first RLOF phase as found in previous BPS studies
\cite{han95a,han95b,han95c,han98,han01}.
We emphasize that because of the wide distribution in $\log a$ it is
almost certain that the resulting orbital period distribution of
post-CE sdB binaries will be wider than what would be obtained in a
more realistic simulation, since in a full BPS simulation only a
subset of this parameter space would be realized by actual systems
(see Paper II). Nevertheless, as we shall show below, this method
still provides a good diagnostic for the CE efficiency.

\subsubsection{Population I BPS simulations}
\begin{figure*}
\epsfig{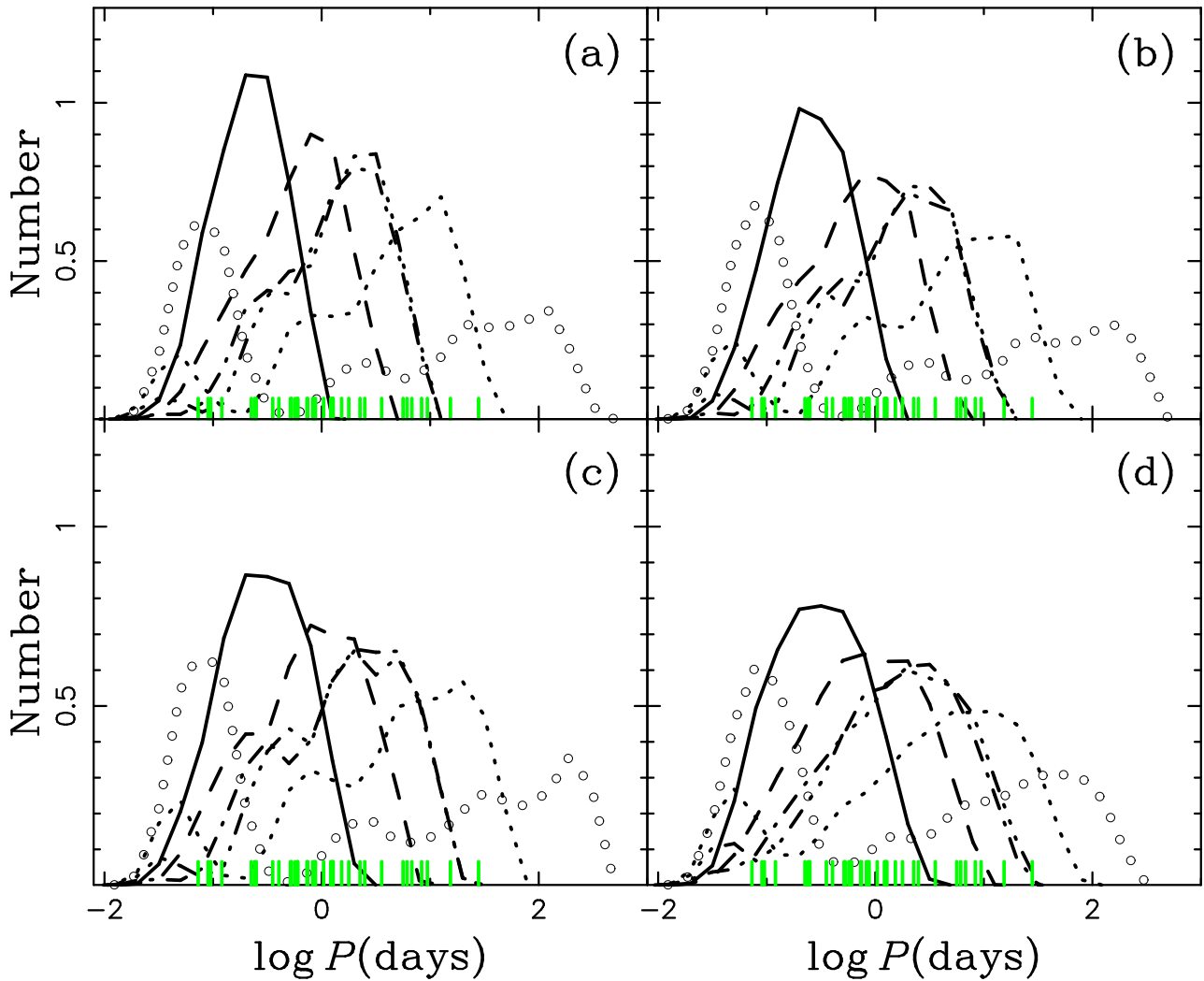}
\caption{
Orbital-period distributions for post-CE sdB stars. Panels (a), (b) and (c)
are for Pop I metallicity, corresponding to stellar models with
Reimers coefficient $\eta=0$, 1/4, 1/2 respectively. Panel (d) is for
$Z=0.004$.
In each panel, the distribution represented by solid, dashed, dot-dashed,
dotted, dot-dot-dot-dashed curves, and circle symbols 
correspond to combinations a) to f) for the CE ejection parameters
$\alpha_{\rm CE}$ and $\alpha_{\rm th}$. Note that the dot-dashed and
dot-dot-dot-dashed curves almost overlap completely.
The short ticks along the period-axis indicate the 
position of observed systems (systems with dM companions, e.g.
PG 1336-018, HW Vir and HS 0705+6700, are not included).
}
\label{sbps-p}
\end{figure*}

\begin{figure}
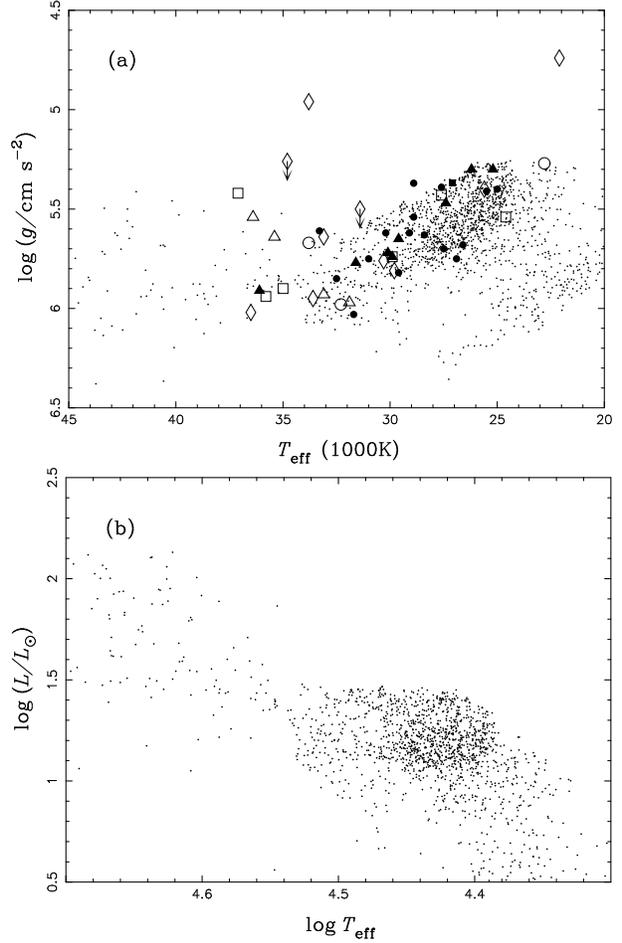

\epsfig{file=sbps-tg.ps,angle=270,width=8cm}
\epsfig{file=sbps-hrd.ps,angle=270,width=8cm}
\caption{
Simulated distribution in the $T_{\rm eff}$ - $\log g$ (top panel) and
H-R diagram (bottom panel) for a `reasonable' model with 
$\alpha_{\rm CE}=\alpha_{\rm th}=0.75$
(dotted curve in panel (b) of
Figure~\ref{sbps-p}). The conversion is based on the sdB evolutionary models
with sdB masses of 0.35 - $0.75M_\odot$ and envelope masses 0.000 - 
$0.006M_\odot$ (see panel (d) of Figure~\ref{sdb-evol}).
}
\label{sbps-tg}
\end{figure}

\begin{figure}
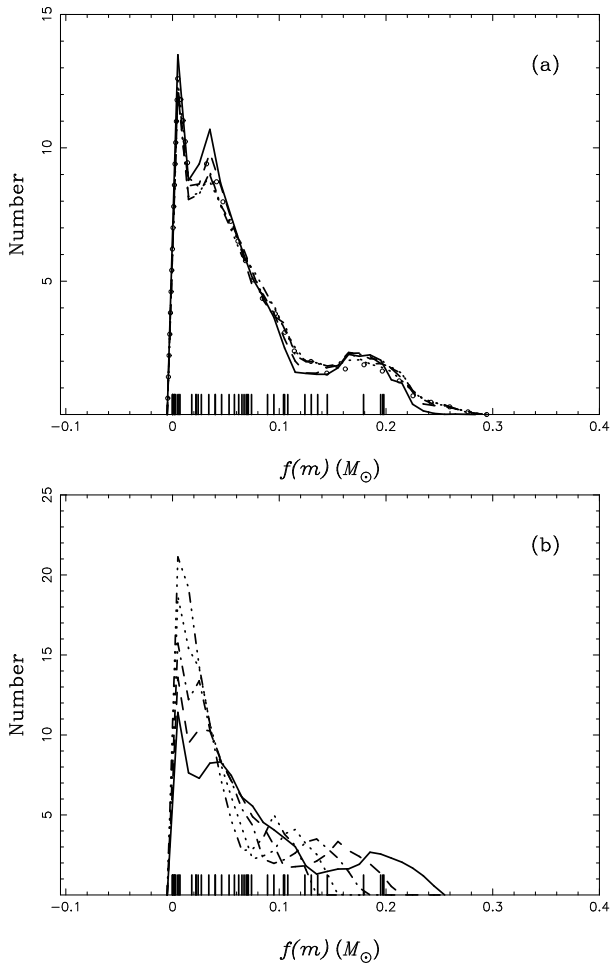

\epsfig{file=sbps-mf1.ps,angle=270,width=8cm}
\epsfig{file=sbps-mf2.ps,angle=270,width=8cm}
\caption{
The distribution of the mass function 
($f(m)={M_2^3\sin^3 i/ (M_1+M_2)^2}$, $M_2$ is the mass of
the WD, $M_1$ is the mass of the sdB star). 
Panel (a)  is for the models in panel (b) of
Figure~\ref{sbps-p} (the different models more-or-less coincide).
The inclination is assumed to be uniformly distributed in solid angle.
The short ticks along the $f(m)$-axis are based on observed systems
(Maxted et al.\ 2001, Morales-Rueda et al.\ 2002).
The distributions generally show three peaks. The first peak
(from the left) is  due to the $\sin^3 i$ factor, the
second peak corresponds to WD masses between 0.55 and $0.65M_\odot$, the third
peak to WD masses between 0.25 and $0.45M_\odot$.
Panel (b) is for the favoured model 
with $\alpha_{\rm CE}=0.75$, $\alpha_{\rm th}=0.75$ and $Z=0.02$
(shown as a dotted curve in panel (b) of Figure~\ref{sbps-p}),
for different masses of the sdB stars (solid, dashed, dot-dashed, dotted,
dot-dot-dot-dashed curves are for
$M_{\rm sdB}$ = 0.4, 0.5, 0.6, 0.7, $0.8M_\odot$, respectively;
$M_{\rm sdB}=0.5M_\odot$ provides an overall best fit to the
observational data points.)
}
\label{sbps-mf}
\end{figure}

\begin{figure}
\epsfig{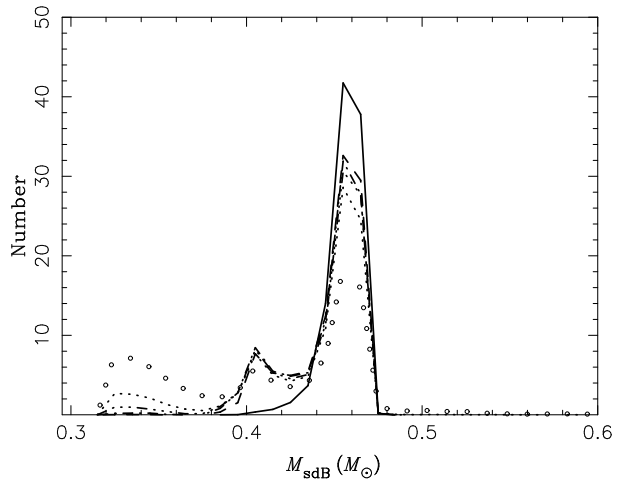}
\caption{
Similar to panel (b) of Figure~\ref{sbps-p}, but
for the distribution of masses of sdB stars.
}
\label{sbps-m}
\end{figure}

In our first series of BPS simulations, we assume a population I metallicity
of $Z=0.02$ and include convective overshooting (of 0.25 pressure scale 
heights). We consider 3 sets of stellar models: (1) models with no stellar 
wind;  (2) models with a Reimers wind mass loss with $\eta = 1/4$; 
and (3) models
with a Reimers wind with $\eta = 1/2$. For each of these 3 sets we take
six combinations of CE parameters 
$\alpha_{\rm CE}$ and $\alpha_{\rm th}$,
including both efficient and inefficient models.
\begin{enumerate}
\item[a)] $\alpha_{\rm CE}=0.2$, $\alpha_{\rm th}=0.0$
\item[b)] $\alpha_{\rm CE}=0.5$, $\alpha_{\rm th}=0.0$
\item[c)] $\alpha_{\rm CE}=0.5$, $\alpha_{\rm th}=0.5$
\item[d)] $\alpha_{\rm CE}=0.75$, $\alpha_{\rm th}=0.75$
\item[e)] $\alpha_{\rm CE}=1.0$, $\alpha_{\rm th}=0.0$
\item[f)] $\alpha_{\rm CE}=1.0$, $\alpha_{\rm th}=1.0$
\end{enumerate}
The resulting post-CE orbital period distributions for these 3 sets are
shown in panels (a), (b) and (c) of Figure~\ref{sbps-p}. 
First, note by comparing the three panels that the
different assumptions about the stellar wind make relatively little difference.
On the other hand, the variation of the CE parameters affects the
orbital period distributions quite dramatically. The overall behaviour
of the various curves is relatively easy to understand.

In the case of inefficient CE ejection, i.e. low $\alpha_{\rm CE}$, the
binary has to spiral in much further during the CE phase before enough
energy has been deposited in the envelope to make ejection possible,
resulting in very short orbital periods. As $\alpha_{\rm CE}$ is
increased, the orbital period distribution shifts towards longer
orbital periods. Increasing the value of $\alpha_{\rm th}$ also increases
the post-CE orbital periods, since it reduces the binding energy of the 
envelope. It also widens the period distribution very substantially. 
This reflects the relative importance of the ionization energy which
varies significantly between a star of $1M_\odot$ and $1.9M_\odot$. 
For a $1M_\odot$ 
star, the total binding energy of the envelope near the tip of the FGB
is close to 0 (in fact, it may be positive, see HPE), implying that the 
envelope is only very loosely bound and that very little spiral-in is
required to eject it. On the other hand, the envelope of a $1.9M_\odot$ star
is much more tightly bound, leading to much smaller post-CE orbital
periods of the systems.
Some of the curves (e.g. the dotted ones) have three peaks. The left
one corresponds to giants with $M_0\ga 2M_\odot$ which have much more
tightly bound envelopes than FGB stars with $M_0\la 1.99M_\odot$. Since
this requires much deeper spiral-in before the envelope can be ejected, 
it leads to the shortest orbital periods. The middle and
the right peaks are the result of CE ejections for FGB stars with
$M_0\la 1.99M_\odot$, where the middle peak corresponds to white dwarf
primaries with $0.25\le M_{\rm WD}\le 0.45M_\odot$ and the right peaks to 
white dwarfs with $0.55\le M_{\rm WD}\le 0.65M_\odot$.
In cases where the CE ejection is not very efficient, i.e.
either $\alpha_{\rm CE}$ or $\alpha_{\rm th}$ is small,
the differences in the period distributions for the two sets of white
dwarfs tends to be small and the latter two peaks tend to merge into one.

The short ticks in these panels along the 
orbital period axis show the periods of sdB binaries with known orbital
periods \cite{max01}. 
As is clear by inspection, the simulated
period distributions taken together cover the whole range of observed
periods, although no single model on its own provides a perfect fit
to the observed distribution. The best models require high values
for both $\alpha_{\rm CE}$ and $\alpha_{\rm th}$
in order to cover the whole range of orbital periods.
Inspection of panels (a), (b) and (c) of Figure~\ref{sbps-p} 
suggests that the dotted simulations
with $\alpha_{\rm CE} = \alpha_{\rm th}= 0.75$ provide the best overall
representation of the observed period distribution. Such a combination
of parameters is physically quite reasonable, since it suggests that
both the ejection of the common envelope and the conversion of thermal (and
ionization) energy are efficient processes, but are not perfect.
This is consistent with theoretical CE ejection simulations \cite{taa00,pod01}
 and independent constraints
from BPS studies of other post-CE binaries with relatively long
orbital periods (e.g. certain symbiotic stars and barium stars;
see Han et al. 1995), which require fairly efficient 
CE ejection. 

In Figure~\ref{sbps-tg}, 
we present various simulated distributions of
the best overall model, i.e. the model with
$\alpha_{\rm CE}=\alpha_{\rm th}=0.75$
and $\eta = 0.25$. 
Panels (a) and (b) of Figure~\ref{sbps-tg} are scatter diagrams
in the $T_{\rm eff}$ -- $\log g$ and the H-R diagram, respectively.
We assumed that the envelope masses are uniformly distributed between
0.0 and $0.006M_\odot$. 
This leads to the concentration of sdB stars
in the upper-right part in panel (a) or the right part in panel (b) 
(see panel (b) of Figure~\ref{sdb-evol}).  
The bulk of the observed distribution of sdB stars in Figure~\ref{sbps-tg}(a)
overlaps nicely with the simulated distribution. The systems that fall
outside the theoretical region are the same as those already discussed
in \S~3.2. These may be more massive sdB stars or those with larger
hydrogen-rich envelopes and may originate from some of the other sdB
channels discussed in subsequent sections. 
There are no observational data points \cite{max01} in the lower-right
part of panel (a) due to observational selection effects (this point will be 
addressed in detail in Paper II). 

Figure~\ref{sbps-mf}a shows the simulated distribution of the mass function 
for sdB stars and Figure~\ref{sbps-mf}b 
compares the mass function distribution for
different masses of the sdB star. While this comparison should not be taken
too literally (since it relies on a very simplified and incomplete
BPS model), it confirms that a model with a sdB mass of $\sim 0.5M_\odot$
provides an overall good fit to the observed distribution, consistent
with earlier findings (e.g. Heber 1986).

Figure~\ref{sbps-m} gives the simulated distributions of masses of sdB
stars.  The distributions have three peaks -- a sharp major peak at
$0.46M_\odot$, a secondary peak at $0.4M_\odot$, and a minor peak at
$0.33M_\odot$. The major peak is caused by systems with a low-mass
ZAMS secondary where the CE is ejected near the tip of FGB (see
Figure~\ref{min-core}).  The secondary peak is due to the fact that
the range in stellar radius which leads to CE ejection near the tip of
the FGB and produces sdB stars is wider for $M_{\rm ZAMS}=1.90M_\odot$
than for $M_{\rm ZAMS}=1.60M_\odot$ (see
Table~\ref{min-core-tab}). Furthermore CE ejection with $M_{\rm
ZAMS}=1.90M_\odot$ results in a low-mass sdB star which has a long
core helium-burning lifetime.  The minor peak contains systems
which had a ZAMS mass greater than the helium flash mass. The figure shows
that almost all the simulated masses of sdB stars are less than
$0.48M_\odot$, which means that the sparsely scattered dots at $T_{\rm
eff} > 35\ 000{\rm K}$ in Figure~\ref{sbps-tg} correspond to the
post-central-helium-burning phase of sdB stars.

\subsubsection{$\boldmath Z = 0.004$ simulations}

In order to test how the results depend on metallicity, we also
performed a series of simulations representing a thick disc population
with a metallicity $Z=0.004$ \cite{gil89}. These results are shown in
panel (d) of Figure~\ref{sbps-p} (for the same combinations of CE
parameters as in \S~3.4.1) and are quite similar to the previous
case. However, the middle and the right peaks
discussed in \S~3.4.1 merge into a single peak.
The reason for this convergence is that
lower-metallicity giants are hotter and more compact, and therefore
have more tightly bound envelopes.
Different WDs then lead to smaller differences between orbital periods 
for the post-CE systems, causing a merging of the two peaks.

\subsection{Discussion}

\begin{figure}
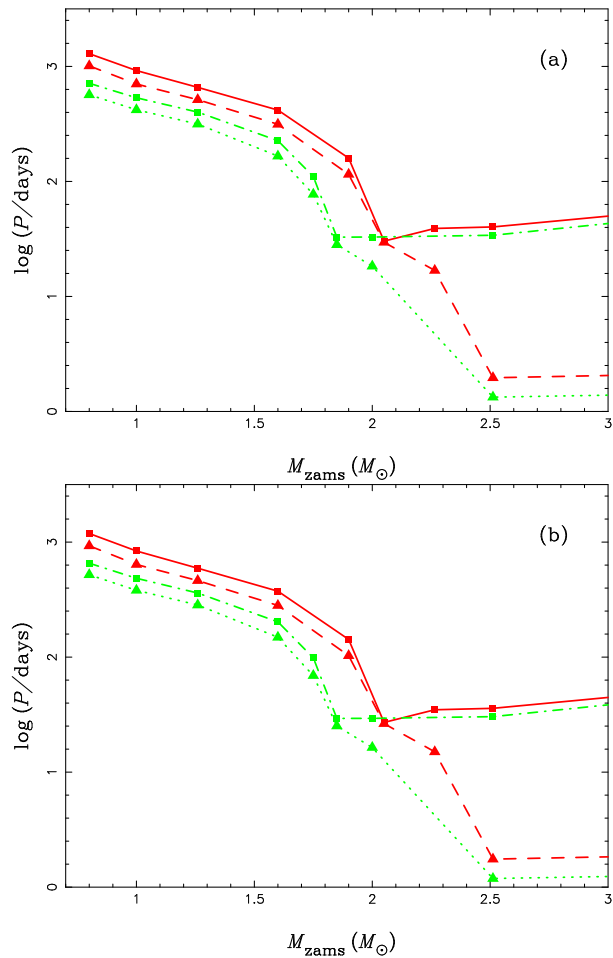

\epsfig{file=p-range1.ps,angle=270,width=8cm}
\epsfig{file=p-range2.ps,angle=270,width=8cm}
\caption{
Panel (a): The orbital period range at the beginning of RLOF
which leads to a spiral-in phase and the formation of a sdB
star in short-period binary
for a WD mass of $0.6M_\odot$ as a function of ZAMS mass. The
dashed and the solid curves are 
for Pop I objects (with overshooting and a 1/4 Reimers' wind). The 
dotted and the dot-dashed
curves are for $Z=0.004$ (with overshooting and a 1/4 Reimers'
wind).
Panel (b): similar to panel (a), but for a WD mass
of $0.3M_\odot$.
}
\label{p-range}
\end{figure}
\begin{figure}
\epsfig{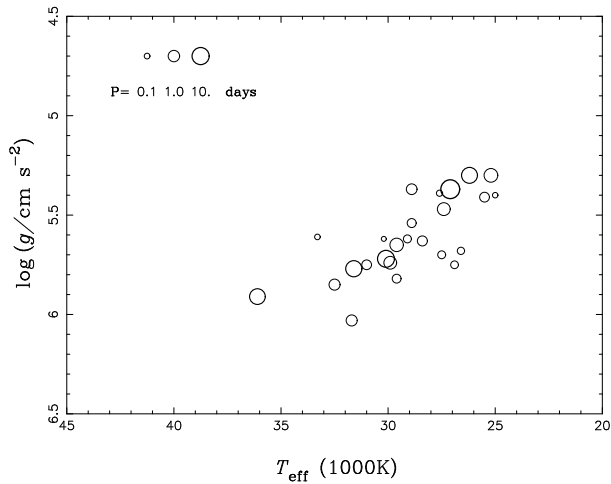}
\caption{
Plot of the distribution of systems with
known orbital periods in the $T_{\rm eff}$ - $\log g$ diagram. Note that
sdB star with long orbital periods appear to have more massive envelopes.
}
\label{p-tg}
\end{figure}
\begin{figure}
\epsfig{file=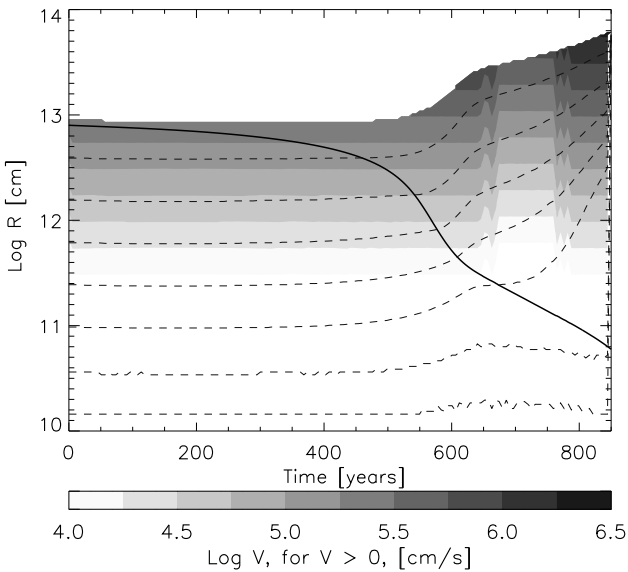,angle=0,width=8cm}
\caption{One-dimensional hydrodynamical calculation simulating
the spiral-in of a $0.3\,M_{\odot}$ compact star inside an evolved
$1.6\,M_{\odot}$ giant, illustrating the delayed dynamical instability that
occurs when the compact objects penetrates into the initially radiative
zone below the convective envelope. The thick solid curve indicates
the assumed location of the spiralling-in object as a function of time;
the dashed curves show the evolution of the radius for various
mass coordinates (from bottom to top: 0.4622, 0.4628, 0.4644, 0.4697,
0.4923, 0.6054, $1.113\,M_{\odot}$). The greyscale indicates the radial
velocity.}
\label{vrp}
\end{figure}
In this section we have shown that helium-burning sdB stars in compact
binaries can be formed if the progenitor giants filled their Roche lobe
when they were close to the tip of the FGB. Figure~\ref{p-range} 
shows the range
of orbital periods at the beginning of mass transfer as a function
of ZAMS mass for which the core (degenerate 
if $M_0\la 1.99M_\odot$ for Pop I 
or $M_0\la 1.8M_\odot$ for $Z=0.004$) will subsequently ignite helium.
Since this is a well-defined evolutionary channel, it provides
an excellent diagnostic to test the criterion for the ejection of the CE.
In Figure~\ref{p-tg} we plot systems with determined orbital periods in the
$T_{\rm eff}$ -- $\log g$ diagram where the size of the symbols indicates
the orbital period. There may be a weak hint that the systems with longer 
orbital periods are less compact and cooler, which implies that they have
bigger envelope masses. This would be consistent with simple expectations
for CE ejection, since larger remnant envelopes may remain bound to the
system if the envelope is ejected at a wider separation.
On the other hand, a small fraction of sdB stars may result from CE 
ejections with $M_0\ga 1.99M_\odot$ for Pop I or
$M_0\ga 1.8M_\odot$ for $Z=0.004$. 
Since the envelope of these more massive giants are more tightly bound 
and the orbital energy available in the spiral-in phase is smaller 
(due to the smaller orbital period of the pre-CE system), sdB
stars produced from more massive giants tend to have very small
orbital periods. Most of these sdB stars have fairly low
masses ($\sim 0.35M_\odot$) and are probably selected against in 
the observational sample (see Paper II).   

Our analysis, however, raises a few further issues. The first is the question
whether the orbital parameters that lead to the formation of the compact
sdB binary, in particular the orbital period range in Figure~\ref{p-range}, 
are being realized by actual binaries
(note that we do not plot the figure for stars more massive 
than $3.0M_\odot$,
as they hardly contribute to the formation of sdB stars due to their
tightly bound envelopes).
This depends on the previous binary
evolutionary phases, which have not been modelled in this paper, but
will be addressed in Paper II.

A second problem is that we found that there is no single combination
of the CE parameters $\alpha_{\rm CE}$ and $\alpha_{\rm th}$ that
produces a perfect fit to the observed period distribution for both
metallicities considered.  It is quite possible that the observed wide
distribution is at least in part a consequence of a large variation of
metallicities and possibly even a variation in the CE ejection
parameters. Considering how uncertain the theoretical modelling of the
CE phase is at present, this would not be very surprising.

Finally we would like to point out an alternative possibility: in
recent studies of the dynamics of the CE ejection process, Ph.P. and
N.I.  (see e.g. the discussion in Podsiadlowski 2001) found that even
in cases where the envelope is not ejected in the initial dynamical
phase, it will always be ejected at a later stage if the spiralling-in
star is relatively compact and penetrates into the initially radiative
layer of the giant.  An example for such a delayed dynamical ejection
is shown in Figure~\ref{vrp}, which represents a 1-dimensional,
hydrodynamical simulation of the spiral-in of a $0.3M_\odot$ compact
star spiralling into the envelope of a 1.6\,$M_\odot$ giant.  The
initial spiral-in is very rapid (see the thick solid curve), but slows
down once the envelope has expanded significantly and the friction
between the spiralling-in binary and the envelope has decreased; in
the subsequent slow spiral-in, the envelope is close to hydrostatic
equilibrium. However, once the compact object reaches the initially
radiative region, a dynamical instability ensues that it is likely to
lead to the ejection of the envelope\footnote{Our calculations were
terminated at this point since they became numerically
unstable.}. While this result should only be considered tentative at
the moment, we note that the conditions for such a delayed dynamical
instability depend mainly on the structure of the giant and cannot be
described, not even in principle, by a simple $\alpha$ ejection
criterion.  The characteristic period obtained in this case is of
order $0.1{\rm d}$, similar to the shortest periods of observed sdB
binaries. If this is the process that leads to the formation of sdB
binaries with the shortest orbital periods, one may expect to see some
structure and possibly even some evidence for a bimodal orbital period
distribution, for which there is no evidence at the moment.

\section{The He WD merger channel}
A second channel that can lead to the formation of a single sdB star
involves the merger of two He white dwarfs \cite{web84,ibe86}.  Close
He WD binaries are formed as a result of one or two CE phases
\cite{web84,ibe85,han95c,ibe97,han98}.  If the orbital period is
sufficiently short (typically less than 6.76 hr for a
$0.3M_\odot$+$0.3M_\odot$ pair to merge in 15Gyrs), gravitational
radiation will cause such a system to shrink until the lighter white
dwarf fills its Roche lobe at a typical period of $\sim 2$\,min.
Mass transfer will be dynamically unstable when the lighter white
dwarf is larger than $\sim$ 2/3 the mass of the more massive
component \cite{pri75,tut79,web84,cam86,web92,han99}.  This leads to
its dynamical disruption and the formation of an accretion disc
surrounding the more massive white dwarf \cite{ben90}.  While the
subsequent evolution has not been modelled in any detail, it is
probably reasonable to expect that a large part of this accretion disc
will be accreted by the more massive component. This accretion will
initially occur on a dynamical timescale, but, as the mass in the disc
decreases and the disc expands, the accretion rate will decrease and
be ultimately determined by the internal viscous processes that govern
the evolution of the disc. As the mass of the white dwarf increases,
there will be a point at which helium is ignited in a shell and the
star will subsequently become a helium core-burning sdB star due to
the inward propagation of these nuclear burning shells
\cite{sai98,sai00}.  Unlike the other scenarios considered in this
paper, this sdB star will be a single object.

\subsection{The conditions for He ignition in He WD mergers}
\begin{table}
 \caption{The minimum merger masses for helium ignition}
 \begin{tabular}{lllll}
 \hline\hline
 $M_1^{\rm He}$ & age & $M_{\rm no-flash}$ & $M_{\rm flash}$ 
   & $M_{\rm ignition}$ \\
 ($M_\odot$) &  (Gyr) & ($M_\odot$) &  ($M_\odot$) & ($M_\odot$) \\
 \hline
  &&&&\\
  0.20 & 0.1 & 0.3860 & 0.3871 & 0.0017\\
  0.20 & 1.0 & 0.3801 & 0.3812 & 0.1860\\
  0.20 & 5.0 & 0.3768 & 0.3778 & 0.2036\\
  &&&&\\
  0.25 & 0.1 & 0.3947 & 0.3958 & 0.2209\\
  0.25 & 1.0 & 0.3822 & 0.3833 & 0.2485\\
  0.25 & 5.0 & 0.3791 & 0.3800 & 0.2537\\
  &&&&\\
  0.30 & 0.1 & 0.4040 & 0.4051 & 0.2969\\
  0.30 & 1.0 & 0.3913 & 0.3923 & 0.3001\\
  0.30 & 5.0 & 0.3908 & 0.3915 & 0.3037\\
  &&&&\\
  0.35 & 0.1 & 0.4234 & 0.4252 & 0.3506\\
  0.35 & 1.0 & 0.4184 & 0.4202 & 0.3541\\
  0.35 & 5.0 & 0.4171 & 0.4186 & 0.3595\\
  &&&&\\
  0.40 & 0.1 & 0.4541 & 0.4555 & 0.4038\\
  0.40 & 1.0 & 0.4516 & 0.4531 & 0.4100\\
  0.40 & 5.0 & 0.4535 & 0.4544 & 0.4133\\
 \hline
 \end{tabular}

 \medskip
  Note - $M_1^{\rm He}$: initial mass of the more massive He WD;
  age: cooling age of the He WD; $M_{\rm no-flash}$: 
  maximum total mass after accretion for which no He flash occurs;
  $M_{\rm flash}$: minimum total mass after
  accretion for which a He flash occurs;
  $M_{\rm ignition}$: ignition point, i.e. the mass coordinate in
  the white dwarf, at which helium ignites.

 \label{ignite-tab}
\end{table}
\begin{figure}
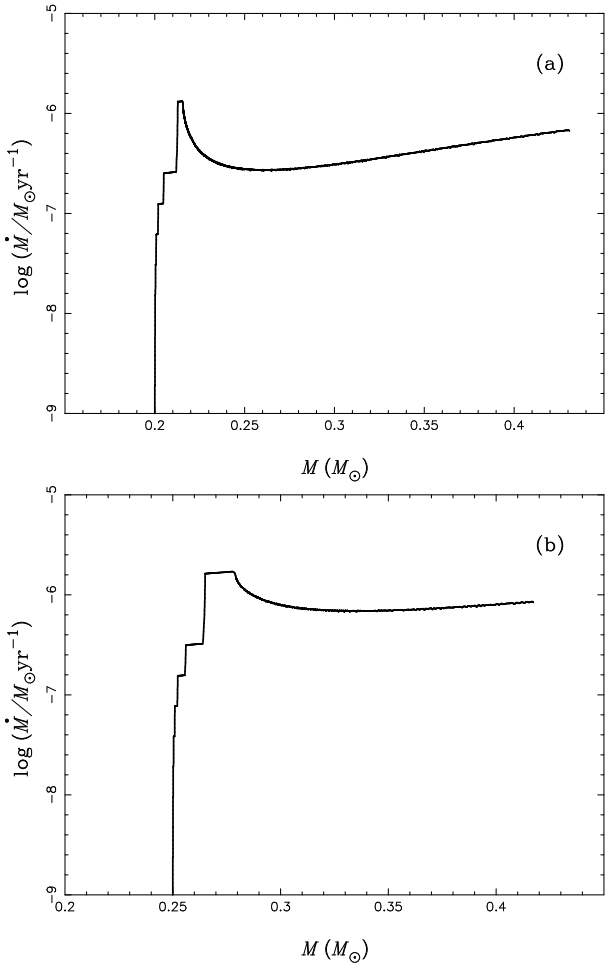

\epsfig{file=accr1.ps,angle=270,width=8cm}
\epsfig{file=accr2.ps,angle=270,width=8cm}
\caption{
Mass accretion rate as a function of total mass 
for initial He WD masses of $0.20M_\odot$ (panel a) and 
of $0.25M_\odot$ (panel b) with cooling
ages of 0.1Gyr accreting at the maximum rate which does not cause 
significant radius expansion.
}
\label{accr}
\end{figure}
\begin{figure}
\epsfig{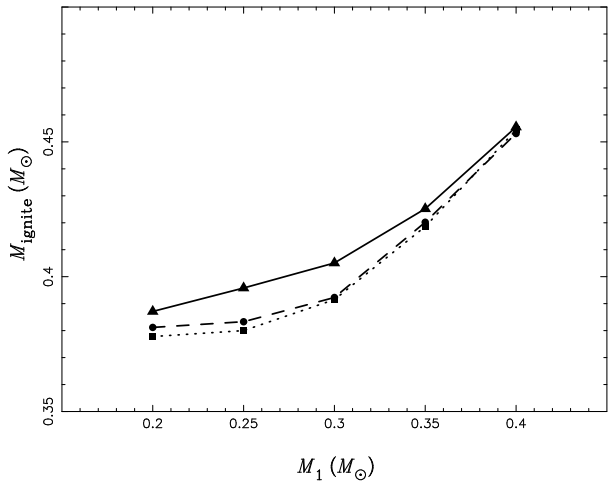}
\caption{
Minimum WD mass at which He is ignited as a function of initial
He WD mass, for different cooling ages (solid curve: 0.1Gyr;
dashed curve: 1.0Gyr; dotted curve: 5Gyr).
}
\label{ignite}
\end{figure}
The conditions for helium ignition depend on the initial mass of the more
massive white dwarf, its initial thermal structure (i.e. its cooling age)
and the accretion history. To determine these, we carried out a
series of detailed accretion calculations where we varied the initial
mass of the massive white dwarf from 0.2 to $0.4M_\odot$ and its initial
cooling age. Specifically, we took the initial WD models from the degenerate
core of a $1M_\odot$ evolutionary calculation and allowed this model to cool
for 0.1, 2 and 5\,Gyr, respectively. At that point we assumed that they
started to accrete pure helium at the maximum rate at which the white dwarf
did not expand drastically (we included the compressional
heating due to the added mass, but not the potential energy of the accreted
matter).
This rate was adjusted continuously, so that
the radius of the accreting white dwarf never exceeded $0.1R_\odot$. 
Panel (a) of Figure~\ref{accr}
presents an example of such an accretion calculation. It shows the mass
accretion rate as a function of the total WD mass for a white dwarf
that initially had a mass of $0.2M_\odot$ and had cooled for 0.1\,Gyr
before the onset of accretion. The critical $\dot{M}$ initially increases
to reach a maximum value of $\sim 10^{-6}M_\odot {\rm yr}^{-1}$, 
which is determined by the maximum radius adopted for
the accreting star. It should be noted that this expansion already
occurs at an accretion rate that is substantially smaller than the 
Eddington-limited accretion rate ($\sim 2\times 10^{-5} M_\odot/{\rm yr}$).
As the white dwarfs continues to accrete, the value of $\dot{M}$ that keeps
the white dwarf at this specified radius decreases first and then rises
because the more massive white dwarfs has a smaller equilibrium radius.
Panel (b) is another example but for an initial He WD mass of  $0.25M_\odot$.
For each combination of initial WD mass and cooling age, we varied 
the amount of 
matter, $\Delta M$, that was accreted and then followed the subsequent
evolution until either helium was ignited or the white dwarf had become
a cool fully degenerate object. In Table~\ref{ignite-tab} 
we list the total mass
of the model that just ignited helium and the model that did not,
respectively, as well as the mass coordinate $M_{\rm ignition}$ 
at which helium ignition occurs 
in the former models. The minimum mass for helium ignition varies
from $\sim 0.38M_\odot$ for the white dwarf with the lowest initial mass
to $\sim 0.45M_\odot$ for the most massive ones (see Figure~\ref{ignite}). This
overall behaviour is determined by the initial thermal structure
of the white dwarf and how it changes as a result of the rapid accretion.
Since the white dwarf becomes more centrally concentrated as its
mass increases and since the accretion timescale is much shorter than
the characteristic cooling timescale, compressional heating will make
the white dwarf less degenerate. This lowers the critical mass for helium
ignition (the minimum mass for helium ignition in non-degenerate
stars is $0.3M_\odot$; e.g. Kippenhahn \& Weigert 1990). Since a white dwarf
of lower mass becomes less degenerate during the accretion phase,
helium is ignited at a lower critical mass.
For each mass, the initially cooler white dwarf can accrete at a higher rate,
hence ignite helium at a lower mass. Note that
ignition generally occurs off-centre, in most cases at a point larger
than the initial mass of the white dwarf (see the last column in 
Table~\ref{ignite-tab}).
We are unable to follow the details of the He flash, but assume that
the nuclear burning front that starts to propagate towards the centre 
after helium ignition will ignite the rest of the core (see the simulations
by Saio \& Nomoto 1998 and Saio \& Jeffery 2000) and that the resulting
core He burning star can be modelled as in \S~3.2.

\subsection{Monte-Carlo simulation of the merger products}
\begin{figure}
\epsfig{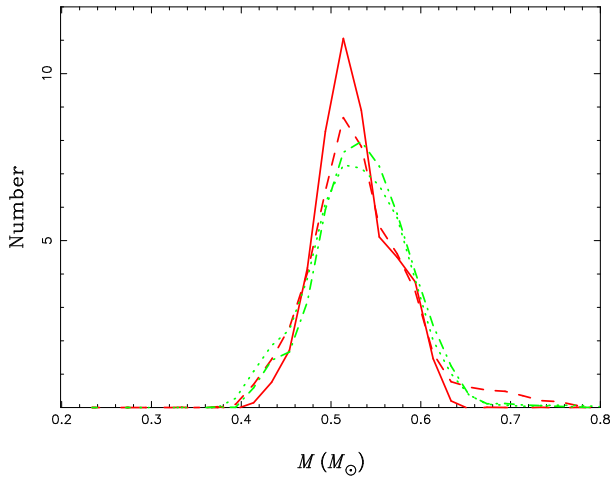}
\caption{
The mass distribution of the He+He WD merger product. Solid, dashed, 
dot-dashed and dotted curves are for simulation sets 1, 2, 3 and 4, 
respectively. 
Simulation set 1 is for 
 $Z=0.02$, $\alpha_{\rm CE}=0.5$, $\alpha_{\rm th}=0.5$;
simulation set 2 is for 
 $Z=0.02$, $\alpha_{\rm CE}=1.0$, $\alpha_{\rm th}=1.0$;
simulation set 3 is for 
$Z=0.004$, $\alpha_{\rm CE}=0.5$, $\alpha_{\rm th}=0.5$;
simulation set 4 is for 
$Z=0.004$, $\alpha_{\rm CE}=1.0$, $\alpha_{\rm th}=1.0$.
}
\label{mgr-m}
\end{figure}
\begin{figure}
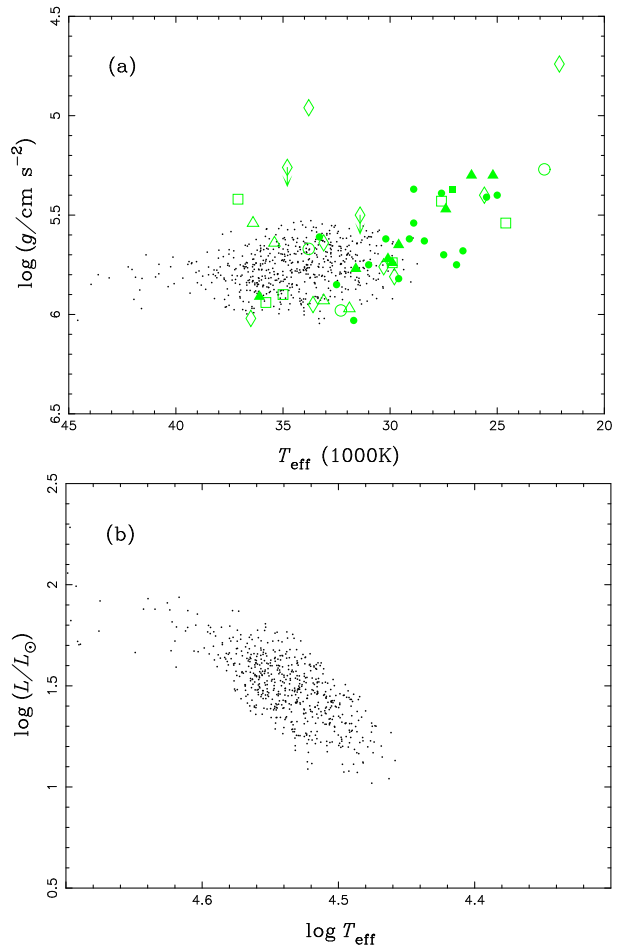

\epsfig{file=mgr-tg.ps,angle=270,width=8cm}
\epsfig{file=mgr-hrd.ps,angle=270,width=8cm}
\caption{
Simulated distribution in the $T_{\rm eff}$ - $\log g$ (panel (a)) and
H-R diagram (panel (b)) for simulation set 1 
($Z=0.02$, $\alpha_{\rm CE}=0.5$, $\alpha_{\rm th}=0.5$).
We assume that the masses of the envelopes of He WD mergers are uniformly
distributed between 0 and $0.001M_\odot$.
}
\label{mgr-tg}
\end{figure}
In Figure~\ref{mgr-m} we
show a simulated distribution of the total mass of the merger
product. These were obtained using the BPS code developed by HPE, which
is further discussed in Paper II, for a fairly standard set of assumptions.
In particular, this simulation assumed a Miller-Scalo 
initial-mass function \cite{mil79}
 for the primary, a flat mass-ratio distribution \cite{maz92,gol94}
and a distribution flat in $\log a$, where $a$ is the orbital separation.
The star-formation rate (SFR) was assumed to be constant with a rate of one
binary with a primary more massive than $0.8M_\odot$ formed per year during
the Galactic lifetime, taken to be 15\,Gyr. We again considered a 
Population I metallicity of $Z=0.02$ and a typical thick-disc metallicity
of $Z=0.004$ (note, however, that in this case the stellar models did 
not include a stellar wind or convective overshooting). For each metallicity
we considered two combinations for the CE efficiencies; a very efficient
model (with $\alpha_{\rm CE}=\alpha_{\rm th}=1$) and a less efficient
model with $\alpha_{\rm CE}=\alpha_{\rm th}=0.5$ (see \S~3.4). 
As Figure~\ref{mgr-m}
shows, the 4 resulting distributions are confined to a relatively narrow
range in mass from $\sim 0.4M_\odot$ to $\sim 0.65M_\odot$ (also see Iben
\& Tutukov 1986). The lower limit is just determined by the minimum mass
for helium ignition (see Table~\ref{ignite-tab}), 
while the upper limit is a consequence
of the previous binary evolution and the timescale for merging 
set by gravitational radiation (see Iben \& Tutukov 1986).
These distributions appear not to be very sensitive to the assumed metallicity,
but seem to depend on the CE ejection parameters. The more efficient the
CE ejection is, the wider the distribution. 
The distribution has a
peak around $0.52M_\odot$.
Panels (a) and (b) of Figure~\ref{mgr-tg} display the
$T_{\rm eff}$ - $\log g$ diagram and HR diagram of the merger products
in these simulations.

\subsection{Discussion}
As Figure~\ref{mgr-m} shows, 
the merger of two He white dwarfs leads to a mass
distribution for the resulting sdB stars that is similar to 
the sdB stars formed
from the CE ejection channel, although this channel also allows the formation
of more massive objects. The birthrates for sdB stars in these simulations
are: $4.6\times 10^{-3}{\rm yr}^{-1}$ (set 1),
$6.5\times 10^{-3}{\rm yr}^{-1}$ (set 2),
$8.3\times 10^{-3}{\rm yr}^{-1}$ (set 3) and 
$1.0\times 10^{-2}{\rm yr}^{-1}$ (set 4),
respectively. These rates are slightly, but not dramatically lower
than the observationally deduced rates of 
$2\times 10^{-14} {\rm pc}^{-3} {\rm yr}^{-1}$ or $0.01{\rm yr}^{-1}$
(by taking an effective Galactic volume of $5\times 10^{11} {\rm pc}^3$)
\cite{heb86}.
Even though these estimates are quite uncertain (the SFR for
 $Z=0.004$  is almost certainly much lower than the SFR for $Z=0.02$),
they still suggest that a significant fraction of sdB stars may form
through this channel. This could help to explain some of the sdB stars
that appear to be more massive than $\sim 0.5M_\odot$ (see \S~3.2), 
provided that these are single objects. Note, however, that to obtain
the distribution in Figure~\ref{mgr-m}, 
we assumed that the mass of the merger
product was the sum of the initial masses of the two He white dwarfs.
This is clearly an upper limit, since some of the mass of the disrupted
lighter white dwarf may remain in a disc around the massive component,
possibly forming asteroids and perhaps even planets in due course
\cite{pod91,liv92}.
The detection of any circumstellar material could potentially
provide an observational test for sdB stars formed through this 
channel. A second uncertainty is related to the amount of hydrogen left
in the merged object. Any hydrogen left from the envelopes of either 
white dwarf component that is mixed with helium and is buried deep
inside the merged object immediately after the merger will ignite
violently and be quickly consumed, altering the thermal structure
of the affected layers in the process (an effect not included in our
accretion calculations). We would therefore generally expect that 
sdB stars from the merger channel have relatively small H-rich 
envelopes and are therefore hotter and more compact than their 
counterparts with more massive H-rich envelopes.

\section{The stable Roche-lobe overflow channel}

A third channel that can produce a sdB star, and in many respects perhaps
the simplest, involves stable mass transfer where a low-mass giant
fills its Roche lobe on the FGB and loses most of its envelope as a result
of stable Roche-lobe overflow (RLOF). Mass transfer stops once the mass
in the H-rich envelope is sufficiently reduced and the radius of the
mass-losing component starts to shrink. If the mass of the degenerate
core is large enough (see \S~3.1), it will still experience a helium
flash and the star may appear as a helium core-burning sdB star in a
binary. Unlike the CE channel, the system will be in a fairly wide 
binary with orbital periods $\ga 1000\,$d (instead of $\la 10\,$d).
This channel has received relatively little attention in the past. This
is at least in part due to a wide-held theoretical misconception concerning
the condition for dynamical mass transfer (see the discussion in 
Podsiadlowski 2001). As is well known, if a fully convective star
(modelled as a polytrope with a polytropic index $n=1.5$) loses mass,
its radius increases, while the Roche-lobe radius decreases if the mass
donor is more massive than the accreting component. This means that the
mass donor  will overfill its Roche lobe by an ever increasing amount,
leading to mass transfer on a dynamical timescale, the formation of
a common envelope and a spiral-in phase (as discussed in \S~3). If mass
transfer is conservative, the critical mass ratio is $\sim 2/3$, i.e.
mass transfer would be dynamically unstable if the mass donor has a mass
larger than 2/3 the mass of the companion star. Since the
minimum mass for a Population I star that can become a giant 
is $\sim 0.9M_\odot$
in a Hubble time, this would imply that the minimum companion mass
has to be larger than $\sim 1.4M_\odot$, i.e. similar to the Chandrasekhar
mass for a white dwarf. Thus, if it were appropriate 
to treat a $\sim 1M_\odot$
giant as a $n=1.5$ polytrope, the parameter space for stable RLOF would be
exceedingly small. However, this argument is not correct for a variety
of reasons. First, it makes several severe simplifications. (1) Giant
stars cannot be modelled as fully convective polytropes, since they have
large degenerate cores. This increases the critical mass ratio for
dynamical mass transfer substantially (Hjellming \& Webbink 1987). (2)
The condition for dynamical instability also depends on the amount of mass
and angular-momentum that is lost from the system (see e.g. Podsiadlowski,
Joss \& Hsu 1992; Han et al.\ 2001; Soberman, Phinney \& van den Heuvel 1997).
(3) Mass loss due to a stellar wind prior to the onset of mass transfer
may significantly reduce the mass of the giant (and increase the
fractional mass of the degenerate core). This mass loss could be significantly
enhanced due to the tidal interaction with the companion \cite{egg89b}. 
A second and perhaps even more fundamental problem with the
simplistic application of such a criterion is that is does not take into 
account the detailed dynamics of the mass-transfer process, in particular
during the turn-on phase in which a substantial amount of mass is already
lost before the dynamical instability occurs. Several recent full
binary evolution calculations have shown that the simplistic criterion
used in most binary BPS studies to date is not really appropriate; e.g. 
Tauris \& Savonije \shortcite{tau99} and Podsiadlowski, Rappaport
\& Pfahl (2002)
have shown
in the case of (sub-)giants transferring mass to a neutron star 
of 1.3/1.4$M_\odot$
that mass transfer is dynamically stable for all giants up to a mass of
$\sim 2M_\odot$ (also see Podsiadlowski et al.\ 1994 for an earlier 
example involving massive stars). On the observational side, it has long
been clear that quite a few systems that should experience dynamical mass 
transfer and a CE phase appear to be able to avoid it (see the discussion
and references in Podsiadlowski et al.\ 1992). In the context of sdB stars,
Green, Liebert \& Saffer \shortcite{gre00} 
have argued strongly that some sdB stars appear
to have companions with large separations. This is consistent with the
findings of Maxted et al.\ (2001), since a fraction of the sdB stars 
in their sample show low radial-velocity variations suggesting that they
are either single or in fairly wide binaries. A detailed reappraisal
of the conditions for dynamical mass transfer is beyond the scope of the
present paper and will be published elsewhere \cite{pod02}.
Here we restrict ourselves to examining the conditions under which stable
mass transfer leads to the formation of a sdB star in a wide binary.

\subsection{The conditions for stable RLOF and the
formation of sdB stars in wide binaries}

\begin{table}
 \caption{Critical masses for stable RLOF}
 \begin{tabular}{lllll}
 \hline\hline
 $M_1^{\rm ZAMS}$ & $M_{\rm c}$ & $M_1^{\rm RLOF}$ & $M_2^{\rm min}$ 
   & $q_{\rm crit}$ \\
 ($M_\odot$) &  ($M_\odot$) &  ($M_\odot$) & ($M_\odot$) \\
 \hline
  0.80 & 0.1992 & 0.7974 & 0.6243 & 1.2773\\
  0.80 & 0.2494 & 0.7945 & 0.6071 & 1.3087\\
  0.80 & 0.2996 & 0.7866 & 0.6071 & 1.2957\\
  0.80 & 0.3501 & 0.7683 & 0.6090 & 1.2616\\
  0.80 & 0.3994 & 0.7328 & 0.5888 & 1.2446\\
  0.80 & 0.4486 & 0.6696 & 0.5225 & 1.2815\\
  &&&&\\
  1.00 & 0.1994 & 0.9975 & 0.8049 & 1.2393\\
  1.00 & 0.2498 & 0.9953 & 0.7898 & 1.2602\\
  1.00 & 0.2992 & 0.9894 & 0.7968 & 1.2417\\
  1.00 & 0.3493 & 0.9759 & 0.8229 & 1.1859\\
  1.00 & 0.3995 & 0.9496 & 0.8354 & 1.1367\\
  1.00 & 0.4482 & 0.9055 & 0.8185 & 1.1063\\
  &&&&\\
  1.26 & 0.1994 & 1.2577 & 1.0339 & 1.2165\\
  1.26 & 0.2493 & 1.2560 & 1.0283 & 1.2214\\
  1.26 & 0.2993 & 1.2515 & 1.0452 & 1.1974\\
  1.26 & 0.3494 & 1.2414 & 1.0804 & 1.1490\\
  1.26 & 0.3985 & 1.2227 & 1.1149 & 1.0967\\
  1.26 & 0.4483 & 1.1905 & 1.1134 & 1.0692\\
  &&&&\\
  1.60 & 0.2483 & 1.5968 & 1.3402 & 1.1915\\
  1.60 & 0.2991 & 1.5941 & 1.3366 & 1.1927\\
  1.60 & 0.3486 & 1.5873 & 1.3943 & 1.1384\\
  1.60 & 0.3981 & 1.5741 & 1.4494 & 1.0860\\
  1.60 & 0.4472 & 1.5522 & 1.4696 & 1.0562\\
  &&&&\\
  1.90 & 0.2495 & 1.8969 & 1.5477 & 1.2256\\
  1.90 & 0.2963 & 1.8959 & 1.6066 & 1.1801\\
  1.90 & 0.3488 & 1.8920 & 1.6366 & 1.1561\\
  1.90 & 0.3982 & 1.8827 & 1.7322 & 1.0869\\
 \hline
 \end{tabular}

 \medskip
  Note - $M_1^{\rm ZAMS}$: ZAMS mass of the primary;
  $M_{\rm c}$: the core mass of the primary at the onset of RLOF; 
  $M_1^{\rm RLOF}$: the surface mass of the primary at the onset of RLOF;
  $M_2^{\rm min}$: minimum mass of the companion (WD/NS) for stable RLOF;
  $q_{\rm crit}$: the critical mass ratio.
 \label{stable-rlof}
\end{table}
\begin{table*}
 \caption{Minimum core mass and orbital 
period for stable RLOF to form a sdB star}
 \begin{tabular}{lllllllll}
 \hline\hline
 $M_1^{\rm ZAMS}$ & $M_1^{\rm RLOF}$ & $M_2$ &
 $M_{\rm c}$ & $\log (P/{\rm d})$ & $q_{\rm RLOF}$ & 
 $\log (P_{\rm sdB}/{\rm d})$ & $M_{\rm sdB}$ & $q_{\rm sdB}$\\
 ($M_\odot$) &  ($M_\odot$) &  ($M_\odot$) & ($M_\odot$) &
 & & & ($M_\odot$) &\\
 \hline
 0.80 & 0.7376 & 0.7263 & 0.3940 & 2.6218 & 1.0537 & 2.9118 & 0.4570 & 0.6526\\
 1.00 & 0.9632 & 0.9023 & 0.3771 & 2.3986 & 1.0822 & 2.9097 & 0.4552 & 0.5112\\
 1.26 & 1.2387 & 1.1310 & 0.3587 & 2.1596 & 1.0962 & 2.9097 & 0.4550 & 0.4022\\
 1.60 & 1.5902 & 1.4556 & 0.3319 & 1.7989 & 1.0967 & 2.8460 & 0.4425 & 0.3049\\
 1.90 & 1.8964 & 1.7234 & 0.2793 & 1.2613 & 1.1006 & 2.6022 & 0.4064 & 0.2354\\
 \hline
 \end{tabular}

 \medskip
  Note - 
  $M_1^{\rm ZAMS}$: ZAMS mass of the primary;
  $M_1^{\rm RLOF}$: the surface mass of the primary at the onset of RLOF;
  $M_2$: the mass of the companion (WD/NS) adopted (close 
  to the minimum mass for stable RLOF according to the results in 
  Table~\ref{stable-rlof}\footnotemark);
  $M_{\rm c}$: the core mass of the primary at the onset of RLOF (RLOF before
      the core mass reaches $M_{\rm c}$ will not result in a helium-burning 
      sdB);
  $P$: the orbital period (in days) at the onset of RLOF;
  $q_{\rm RLOF}$: the mass ratio at the onset of RLOF; 
  $P_{\rm sdB}$: the orbital period of the sdB star formed;
  $M_{\rm sdB}$: the mass of the sdB star formed; 
  $q_{\rm sdB}$: the mass ratio of the sdB binary.
 \label{stable-rlof-sdb}
\end{table*}
\footnotetext{This makes  
  the WD mass as small as possible, or as close as possible to a typical
  WD mass of $0.6M_\odot$. Note, however, that, for a higher WD mass,
  RLOF is more stable and the mass transfer-rate is lower. As a consequence 
  the core can grow more massive making He ignition more likely. In this
  case, the minimum orbital period $P$ for the formation of a sdB star
  could be lower than indicated in the table.}
\begin{figure}
\epsfig{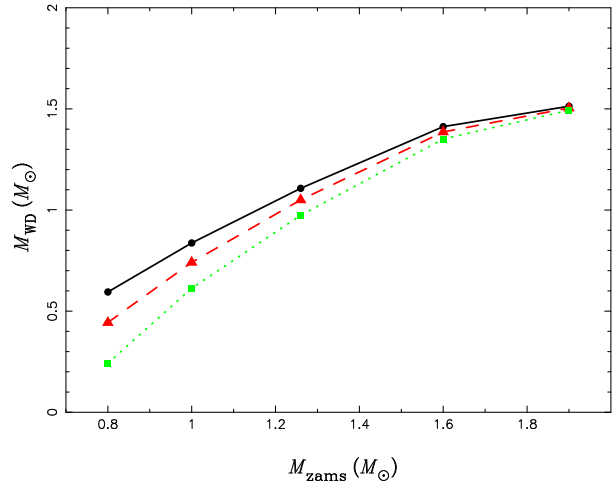}
\caption{
Minimum WD mass for stable RLOF (see Tables~\ref{stable-rlof} \&
~\ref{stable-rlof-sdb}) as a function of ZAMS mass that leads to the
formation of a sdB star. The solid curve is for a 1/4 Reimers' wind,
dashed and dotted curves are for Reimers' wind mass-loss rate with
$\eta=1$ and $\eta=2$ (i.e. significantly enhanced wind-loss rate;
e.g. due to tidal interactions).}
\label{min-wd}
\end{figure}
To determine the conditions for the formation of sdB stars through the
stable RLOF channel, we performed a series of binary stellar evolution
calculations for mass donors with different masses on the zero-age
main sequence (ZAMS), ranging from 0.8 to $1.9M_\odot$. For each mass, we
systematically varied the mass of the core at the beginning of mass
transfer and the mass of the companion star. In these calculations we
assumed that mass transfer was completely non-conservative, and that
all the mass that was lost from the system carried with it the orbital
angular momentum of the accreting component (as appears to be most
appropriate if the accretor is a white dwarf). In the standard set of
calculations, we included a Reimers-type wind with $\eta=1/4$ (see
equ.~\ref{reimers}) before the mass-transfer phase. We switched this wind off
once the mass-transfer rate exceeded the value given by equ.~\ref{reimers} by
 a factor of 100 and did not include stellar-wind mass loss after the end
of the mass-transfer phase. In each calculation, we checked first
whether mass transfer was dynamically stable. In cases, where mass
transfer is dynamically unstable, there is no solution for the
mass-transfer rate, $\dot{M}$, for which the radius of the secondary
can be equal to the Roche-lobe radius (see Han, Tout \& Eggleton 2000
for the treatment of
the surface boundary condition). If mass transfer is stable, we
continued mass transfer until the mass donor started to shrink below
its Roche lobe, terminating the mass-transfer phase.  If, at this
point, the mass of the H-exhausted core exceeded the appropriate
minimum core mass for subsequent helium ignition (as determined in
\S~3.1\footnote{In a number of cases, we continued the calculations
up to the point of helium ignition to confirm that the limits
determined in \S~3.1 are also applicable in this case.}), it may
appear as a helium core-burning sdB star.

The results of these calculations are summarized in 
Tables~\ref{stable-rlof} and ~\ref{stable-rlof-sdb}, 
which show for each ZAMS mass the core mass and the total mass of the
mass donor at the beginning of mass transfer and the minimum mass of
the secondary (and critical mass ratio) for which mass transfer is
stable and leads to the formation of a helium-burning sdB star.  These
results demonstrate, as discussed above, that mass transfer is
dynamically stable even if the mass donor is substantially more
massive than the secondary and that sdB stars can form through this
channel without any non-standard assumptions. For stars with ZAMS
masses $\ge 1.6M_\odot$, the secondary mass has to be larger than
$1.34M_\odot$, which is similar to the maximum mass of a white dwarf, and
hence does not correspond to realistic systems with white-dwarf
accretors. 
Note that, for each mass, the critical mass ratio tends to
decrease for the initially more evolved systems, since for these the
evolutionary timescale is shorter and hence the mass-transfer rate
higher than for the less-evolved ones.  This has the consequence that
the core mass grows less during the mass-transfer phase. 
One may also notice that some of
the behaviour in Table~\ref{stable-rlof} is non-monotonic. 
The non-monotonic behaviour of $q_{\rm crit}$ at the lowest $M_{\rm c}$
is caused by primaries with a core mass $M_{\rm c}$ near the base
of the FGB, where the core is not very degenerate and the envelope is not yet
fully convective. 
The non-monotonic behaviour
for $0.8M_\odot$ stars with the largest $M_{\rm c}$ is a consequence
of its thin envelope mass.

While these calculations show that sdB stars can form in wide binaries
without any non-standard assumptions, it is quite plausible, perhaps
even likely, that the wind mass-loss rate before the beginning of the
RLOF phase will be enhanced due to the tidal interaction with the
companion (as originally proposed by Eggleton \& Tout 1989). The main effect
this has is to increase the overall parameter space for which this
channel produces sdB stars in wide binaries. This is illustrated in
Figure~\ref{min-wd}, which shows the minimum mass of the accreting white dwarf
as a function of the ZAMS mass of the donor for the standard model
used above and
two sequences of calculations, where we assumed that the pre-RLOF wind
mass loss rate was enhanced by a factor of 4 and 8, respectively
(i.e. we took $\eta = 1$ and 2).

\subsection{Binary Calculations}
\begin{figure}
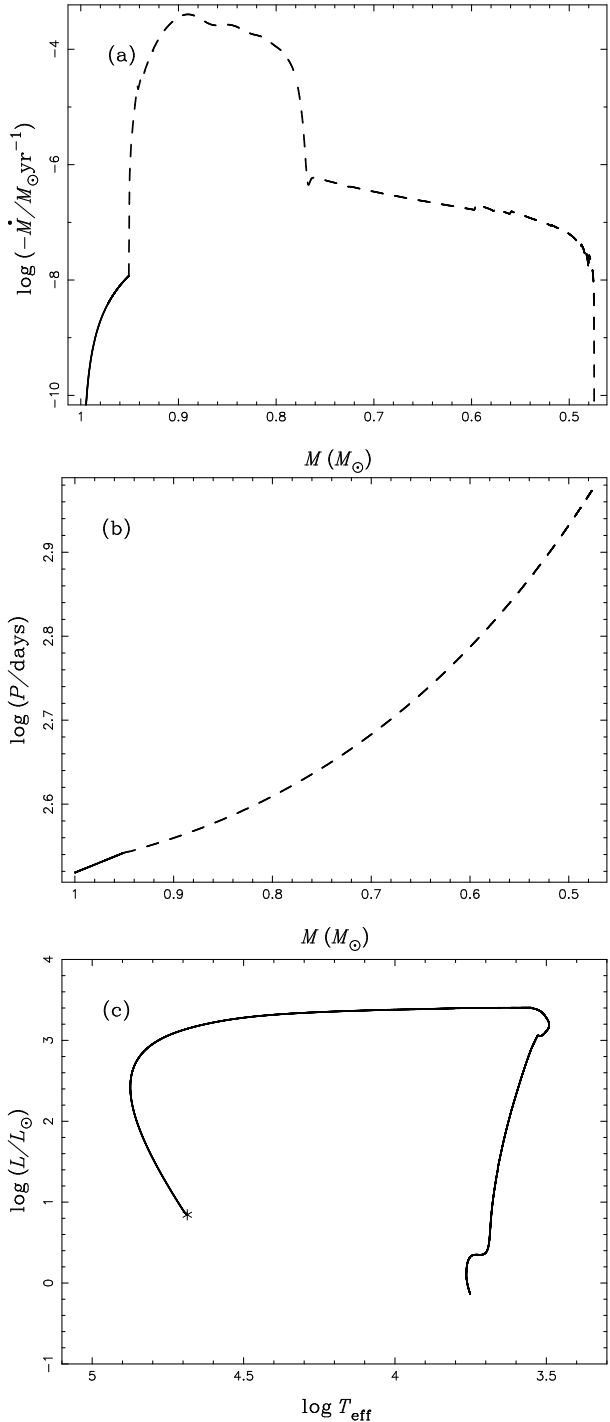

\epsfig{file=binary1.ps,angle=270,width=8cm}
\epsfig{file=binary2.ps,angle=270,width=8cm}
\epsfig{file=binary3.ps,angle=270,width=8cm}
\caption{
Evolution of mass-transfer rate (panel a), orbital period (panel b) as a 
function of mass, 
evolutionary track in the H-R diagram (panel c) to demonstrate
the case of stable RLOF for a binary with a giant donor with an initial
mass of $1M_\odot$ and a $0.84M_\odot$ WD companion (Pop I, with overshooting,
1/4 Reimers' wind). The solid curves in panels a and b show the evolution
before the onset of RLOF, i.e. 
due to a stellar wind. No stellar wind was included during and after RLOF. 
}
\label{binary}
\end{figure}
Figure~\ref{binary} shows a representative binary calculation from one of the
sequences in the previous section for a star with a ZAMS mass of
$1M_\odot$. At the beginning of mass transfer, the mass donor has a core
mass of $0.3975M_\odot$ and a total mass of $0.9508M_\odot$, and the mass of
the companion star is $0.84M_\odot$. With these parameters, mass transfer
starts at an orbital period of 348.4\,d. Initially, mass transfer
occurs on a thermal timescale and reaches a maximum of $\sim 4\times
10^{-4}M_\odot {\rm yr}^{-1}$. After the mass ratio has been
reversed and the star has regained thermal equilibrium, mass transfer
settles to a rate of $\sim 4\times 10^{-7}M_\odot {\rm yr}^{-1}$ 
and gradually decreases as the secondary ascends the giant
branch. Once the mass in the H-rich envelope drops below $0.021M_\odot$, 
the secondary shrinks below the Roche lobe
and mass transfer stops. As the remnant envelope collapses, the
secondary quickly moves across the H-R diagram and ultimately becomes
a sdB star of $0.4745M_\odot$ in a wide binary with an orbital period of
948.9\,d.

Whether the secondary becomes a sdB star and its location both in the
$T_{\rm eff}$ -- $\log g$ and the H-R diagram also depends on mass loss
via a stellar wind after the RLOF phase.  For example, consider a
binary with an orbital period of 606\,d consisting of a giant with a
total mass of $0.9206M_\odot$ and a core mass of $0.4343M_\odot$ (which
corresponds to one of the sequences with a $1M_\odot$ ZAMS star) in orbit
with a $0.82M_\odot$ white dwarf. If wind mass loss is switched off once
the system has entered the RLOF phase, the giant starts to ignite
helium when its core mass reaches $0.4731M_\odot$, but its total mass is
still $0.5210M_\odot$ (the system has an orbital period of 1306\,d at this
point). If there is no further mass loss, the secondary will then
settle on the normal horizontal branch, burn helium in the core and
then ascend the asymptotic-giant branch (AGB) (this typically requires
an envelope mass larger than $\sim 0.05M_\odot$; see Dorman, Rood,
O'Connell 1993).  
On the other hand, if wind mass loss is not switched off during the RLOF phase
but continues at a rate of 1/4 of the Reimers rate, the total mass is
$0.4864M_\odot$ when helium is ignited in the core of mass
$0.4731M_\odot$. In this case it has the appearance of a sdB star lying in
the upper-right corner of the $T_{\rm eff}$ -- $\log g$ diagram.

\subsection{Discussion}
As we have shown in this section, stable RLOF provides a third channel
for the formation of sdB stars. These will generally be in wide
binaries with typical orbital periods of $400-1500\,$d.
The mass distribution is similar to the distribution in
the CE channel.  However, one might expect that the envelope masses
could be systematically larger than in the CE ejection case, since the
orbital period is much longer, which means that a larger envelope mass
may remain bound to the degenerate core when the donor becomes
detached. This would suggest that sdB stars formed through this
channel would be less compact and cooler than their counterparts from
the CE channel.

The importance of this channel is difficult to assess.  While standard
assumptions lead to the formation of sdB stars, the allowed range for
the binary parameters is probably quite small in this model, since it
requires relatively massive white dwarfs, which are not very common.
However, this parameter range could be dramatically increased if the
stellar-wind mass-loss rate is significantly enhanced, e.g. due to the
tidal interaction with the companion star \cite{tou88}. On the other
hand, observational surveys could in principle allow the calibration
of such enhanced stellar winds.

One may notice that we only deal with WD companions in this section,
though stable RLOF with MS companions can also produce sdB stars.
Observational surveys show that at least half of all sdB stars have
cool MS companions (see Section 1). This seems to be in contradiction
with the observations by Maxted et al.\ \shortcite{max01} that the 
majority of sdB stars are short-period binaries with WD companions,
as it implies a binary fraction larger than 1. The contradiction
is due to various observational selection effects and will be addressed in
Paper II. 


In this section, we made the simplifying assumption that mass transfer
to a WD companion is completely non-conservative in the stable RLOF
channel.  This assumption is appropriate for low mass-transfer rates,
where nova explosions are believed to be effective in expelling all of
the transferred matter, but probably not for higher rates (larger than
$\sim 10^{-7}M_\odot\ {\rm yr}^{-1}$), where the white dwarf may be
able to accrete most of the transferred matter and burn it steadily
(as in supersoft X-ray binaries). On the other hand, for even higher
rates the white dwarf will start to swell up
\cite{nom79} and may then lose most of the transferred mass again, possibly in
the form of an optically thick wind \cite{hac96}. Clearly all of these effects
need to be studied further and should be included in future studies.

\section{Summary}

In this paper, we have demonstrated that the three binary evolution
channels that have been proposed for the formation of
sdB (and related sdO/sdOB) stars may all contribute to the observed
population.

In the CE ejection channel, which may account for more than 2/3 of all
sdB stars, dynamically unstable mass transfer near the tip of the FGB 
results in the formation of a CE and spiral-in phase, leaving a short-period
binary after the envelope has been ejected. The system becomes
a sdB binary if helium is ignited. Using detailed stellar evolution 
calculations, we have determined how close to the tip of the FGB the progenitor
has to be at the onset of RLOF. Using simplified binary population synthesis
calculations, we have been able to show that the CE ejection process
has to be very efficient and that the ionization energy in the envelope
has to be included in the ejection criterion in order to be able to explain 
the observed orbital period distribution.

In the stable RLOF channel, the progenitor systems experiences stable mass 
transfer in which the giant is stripped off its envelope as a result
of the mass transfer. If this occurs near the tip of the FGB, the remnant
helium core will still ignite helium in the core and become
a sdB star in a binary with a long orbital period and a fairly thick
hydrogen-rich envelope as compared to the other channels. Using detailed
binary evolution calculations, we demonstrated that this channel usually
requires a fairly massive white dwarf companion or enhanced stellar
wind mass loss before the onset of RLOF (e.g. tidally enhanced winds).

Double He WDs may coalesce due to gravitational wave radiation. 
When helium is ignited in the merger, a single sdB star is formed, and
its hydrogen envelope is likely to be very thin. We have determined
the conditions for which the merged system will be able to ignite helium.

In a follow-up paper we will implement these results in full binary
population synthesis calculations to assess their relative importance
and to allow direct comparison with observed subdwarf populations.

\section*{Acknowledgements}

We thank Drs. A. Lynas-Gray and S. Yi for many stimulating discussions.
We are grateful to Dr. O. Pols, the referee, for his valuable suggestions
which helped to improve the paper.
ZH thanks the Department of Astrophysics, Oxford for its hospitality.
This work was in part supported by a Royal Society UK-China Joint Project
Grant (Ph.P and Z.H.), the Chinese National Science Foundation under 
Grant No.\ 19925312, 10073009 and NKBRSF No. 19990754 (Z.H.).

\end{document}